\documentclass[fleqn,usenatbib]{mnras}

\usepackage{newtxtext,newtxmath}

\usepackage[T1]{fontenc}

\DeclareRobustCommand{\VAN}[3]{#2}
\let\VANthebibliography\thebibliography
\def\thebibliography{\DeclareRobustCommand{\VAN}[3]{##3}\VANthebibliography}

\usepackage{graphicx}	
\usepackage{amsmath}	
\usepackage{mathtools}

\title[Semi-confined SNe in H {\scshape ii} regions]{Semi-confined supernova feedback in H {\scshape ii} region bubbles}

\author[C. S. C. Lau \& I. A. Bonnell]{
Cheryl S. C. Lau$^{1}$\thanks{E-mail: cscl1@st-andrews.ac.uk (CSCL)}
and Ian A. Bonnell$^{1}$
\\
$^{1}$School of Physics and Astronomy, University of St Andrews, St Andrews KY16 9SS, UK
}

\date{Accepted XXX. Received YYY; in original form ZZZ}

\pubyear{2025}

\begin{document}
\label{firstpage}
\pagerange{\pageref{firstpage}--\pageref{lastpage}}
\maketitle

\begin{abstract}
Galactic-scale simulations rely on sub-grid models to provide prescriptions for the coupling between supernova (SN) feedback and the interstellar medium (ISM). Many of these models are computed in 1-D to allow for an efficient way to account for the variability of properties of their local environment. However, small-scale simulations revealed that the release of energy from SNe within molecular clouds can be highly asymmetrical. This is largely due to the presence of pre-SN feedback, such as ionizing radiation, that are able to carve cavities and channels around the progenitors prior to their detonation. Being partially confined, the SN energy escapes into the outer ISM preferentially through these channels, departing from the spherically symmetric 1-D descriptions. To understand by how much the feedback output could differ, we present a theoretical model for a semi-confined SN. The problem concerns a SN expanding into an evolved H {\scshape ii} region, bounded by a molecular cloud with pre-existing vents. With the aid of simple 3-D hydrodynamical simulations, we show that this mode of energy release increases the local dynamical impact of the outflows, and extends the timescales over which the SN is energetically coupled to the surrounding matter. We also show that the amount of small-scale solenoidal turbulence driven by semi-confined SNe may be amplified. 
\end{abstract}

\begin{keywords}
ISM: supernova remnants -- ISM: bubbles --  HII regions -- methods: analytical -- methods: numerical
\end{keywords}



\section{Introduction}

From the scales of molecular clouds up to galaxies, it is generally recognized that supernova (SN) feedback is one of the key mechanisms for regulating star formation \citep[e.g.][]{larson74,hopkins11,kortgen16}. The energy imparted on to the interstellar medium (ISM) from Type II SNe, on Galactic average, dominates over the contributions from ionizing radiation and stellar winds over the lifetime of the OB-type stars \citep[][]{maclowklessen04}. SNe are also largely responsible for cosmic ray acceleration \citep[e.g.][]{pfrommer17}, chemical enrichment of the ISM \citep[e.g.][]{goswami21}, and the replenishing of supersonic turbulence in molecular clouds \citep[e.g.][]{joungmaclow06,hennebelleiffrig14}. SN explosions from multiple massive stars can collectively form supperbubbles that drive the galactic fountains \citep[e.g.][]{keller15,kim17}, and play a fundamental role in determining the star formation efficiencies across all scales \citep[e.g.][]{ostrikershetty11}. It is therefore inevitable to investigate the energy and momentum deposition from SNe on its local environment in order to understand how these ecosystems operate. 

Quantifying the coupling efficiency between SNe and their immediate surroundings remains a challenge. For one, SN shocks are subjected to radiative cooling. Depending on the density structure of the molecular cloud, a significant portion of the injected thermal energy could be radiated away before making any substantial dynamical impact on the cold ISM \citep[e.g.][]{mckeeostriker77,cowie81,walchnaab15}. For two, which is often deemed the primary reason, is that the evolution of SN remnant is heavily influenced by the pre-SN stellar feedback \citep[e.g.][]{borkowski96,dwarkadas05,rogerspittard14,lucas20}. This feedback could originate from the SN progenitor star itself during its earlier phases, or from the neighbouring massive stars in the OB association \citep[][]{oeyclarke97}. Either way, SN remnants likely expand into a circumstellar environment that has been modified. 

Ionizing radiation, for example, are shown to be able to create cavities and bubbles of around $10\ \mathrm{pc}$ in radius inside molecular clouds \citep[e.g.][]{dale13c,dale14,fichtner24}. The Lyman continuum radiation from massive stars ionizes the surrounding gas up to the Str\"{o}mgren radius \citep[][]{stromgren39} and heats them to approximately $10^4\ \mathrm{K}$ with the excess energy \citep[][]{osterbrock74}. Upon reaching ionization equilibrium, the H {\scshape ii} region begins to expand under thermal overpressure \citep[D-type expansion; ][]{spitzer78}. Because the ionization front is supersonic to the neutral ambient medium, a dense shell of shocked neutral gas is formed, sweeping across the ISM as the H {\scshape ii} region evolves \citep[e.g.][]{hosokawainutsuka06,raga12b,bisbas15}. 

Similarly, stellar winds from Wolf-Rayet stars with terminal velocities up to $3000\ \mathrm{km\ s^{-1}}$ are also capable of driving a circumstellar shell of shock-heated gas \citep[e.g.][]{castor75,weaver77,geen22}. This also applies to the massive runaway stars that have migrated \citep[][]{haid18,meyer24}. Indeed, stars that pass frequently through the dense regions are likely unable to develop wind-blown bubbles, given that they may also be inhibited by the large-scale accretion flows \citep[][]{dalebonnell08}. Yet, a later study by \citet{mackey14} proposed that static shells may still be formed around a runaway progenitor if the cavity is subjected to external ionizing radiation. This illustrates the necessity to consider SN shock collisions with the surrounding swept-up matter \citep[][]{dwarkadas05,walchnaab15}, whilst also accounting for the non-linear interplay between the winds and the ionizing radiation \citep[e.g.][]{freyer06,haid18,geen23}. 

However, galactic-scale simulations are often unable to resolve down to parsec or sub-parsec scales where these feedback interactions take place. The lack of resolution may also provoke numerical overcooling \citep[e.g.][]{katz92,springelhernquist02,creasey11,dallavecchiaschaye12} where radiative losses are erroneously amplified. For this reason, they rely on sub-grid models, that serve as prescriptions of the feedback’s overall impact \citep[e.g.][]{thackercouchman00,stinson06,oku22}. The use of sub-grid models addresses the spurious cooling and allows for a more accurate estimate of the energy deposition. Nonetheless, the precise injection method and the SN energy budget can introduce large uncertainties to the galactic evolutionary models \citep[e.g.][]{rosdahl17,kellerkruijssen22}. Refining these feedback prescriptions has thus been a priority in numerical development, and the results from cloud-scale simulations of SNe \citep[e.g.][]{kortgen16} provide this crucial missing piece of puzzle. 

To study the local energy deposition from SN, recently there has also been a growing tendency to employ 1-D hydrodynamic models. Their low computational costs enable large parameter grids to be scanned to account for a variety of complex ISM environments, such as different density distributions or turbulent structures \citep[][]{haid16}. It may also account for different evolutionary phases of the feedback bubble from varying star formation efficiencies \citep[][]{rahner17}. These models can further cover a range of SN progenitors by adopting stellar models, which allow the ejecta energy to be self-consistently computed from the stellar core properties \citep[][]{fichtner22,fichtner24}. 

An issue with 1-D models, however, is that they are bound to model the SN output in a spherically symmetric manner\footnote{For 1-D models, statistical methods may be applied to account for any directional variations \citep[e.g.][for densities in turbulent medium]{haid16}, or through parametrization \citep[e.g.][for shell cover fractions]{harperclarkemurray09,rahner17}, but it would still require results from 3-D simulations to formulate these compensation methods.}. Previous studies by \citet{rogerspittard14,wareing16,wareing17,lucas20} unanimously suggested that this is often not the case. The pre-SN feedback, including ionizing radiation and stellar winds, are able to follow the paths of least resistance within the molecular cloud where gas densities are relatively low. These \textit{channels} serve as chimneys that connect the embedded feedback-driven cavities towards the outer ISM. The energy from the first SN, which detonates a few Myr after the onset of feedback, would preferentially leave the cloud through the pre-existing channels, leaving a minimal impact on the dense structures \citep[][]{lucas20}. As such, the outflows from SNe exploding within feedback-driven bubbles are likely highly anisotropic. 

In fact, this mode of SN energy release has been long theorized by \citet{morfilltenoriotagle83,fallegarlick82,tenoriotagle85}, who applied 2-D hydrodynamical calculations to model the SN energy outbreaks from embedded progenitors near the edge of molecular clouds (see also H {\scshape ii} region \textit{champagne flows}). These models were proposed to explain the the complex velocity profile of the Cygnus Loop. The large asymmetry in its blowouts led to the conjecture that its peculiar morphology is caused by SN shocks encountering the inner walls of a wind-blown bubble, with breakouts through the shell into lower density ISM \citep[][]{aschenbachleahy99,uchida08}. Various alternative interpretations have been proposed \citep[e.g.][]{meyer15,fang16} but nevertheless support the idea that a SN remnant evolving within a pre-existing non-spherical cavity shaped by its progenitor is responsible for the observed mixed-velocity emissions \citep[see also e.g.][]{borkowski17}. 

The question here is: by how much would the energy and momentum deposition differ if the SNe were subjected to partial confinement? In another words, to what extent do 1-D sub-grid models hold given that SN outflows are likely expelled in preferential directions? Past studies that investigated the porosity of feedback bubbles typically place their focus on the \textit{loss} of kinetic energy \citep[][]{rosen14}, momentum flux \citep[][]{dalebonnell08} or H {\scshape ii} gas pressure \citep[][]{harperclarkemurray09,lopez11,lopez14} in order to explain the inconsistencies between the feedback energy input and the observed low X-ray emission luminosities. Less attention was given to their \textit{positive} impacts. \citet{geen23}, for example, suggested that the plumes formed by wind bubbles that break out from thin chimneys can lead to faster expansion velocities, and drive stronger fluid eddies at where the plume interacts with the ISM. This motivates our paper, to investigate what physical properties of the feedback outflows are \textit{enhanced} due to the semi-confinement. 

This paper is organized as follows. In Section~\ref{sec:analytical_models}, we present our novel analytical model of a partially confined SN. In Section~\ref{sec:numerical_methods}, we describe the numerical methods to test these theoretical models with 3-D hydrodynamical simulations. Section~\ref{sec:results} presents the results examining the outflows, the energy and momentum deposition, and the driven turbulence from semi-confined SNe. Finally, we summarize in Section~\ref{sec:conclusion}.

\section{Theoretical model} \label{sec:analytical_models}

We first tackle our problem via an analytic approach to model how SN energy escapes the feedback bubbles. In the present study, we focus on H {\scshape ii} regions driven by photoionization, since their interior properties are relatively more uniform and hence allow for simpler approximations. Extensions to treat wind-blown bubbles or a combination of both would be ideal in future work. 

\begin{figure*}
    \centering
    \includegraphics[width=6.5in]{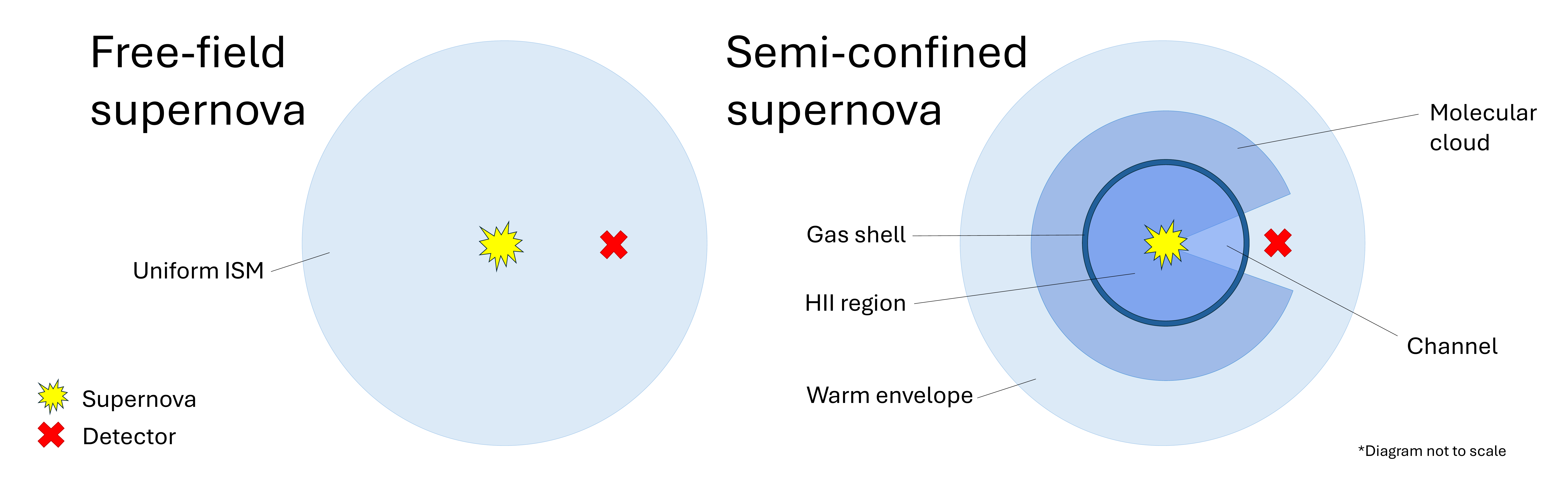}
    \caption{\textit{Left}: Illustration of a free-field SN model. The SN detonates within a uniform medium, and a detector (marked by a cross) is placed at some distance away from the progenitor to measure the gas properties (see Section~\ref{sec:outflow}). \textit{Right}: Illustration of a semi-confined SN model. The progenitor is embedded inside an H {\scshape ii} region bordered by a dense shell of swept-up gas, surrounded by a molecular cloud. The cloud is enveloped by warm diffuse gas that represents the outer ISM environment. The H {\scshape ii} region and the warm envelope are connected by a low-density channel. A detector is placed at the channel opening to measure the local perturbation caused by the outflow (see Section~\ref{sec:outflow}). A confined SN model is identical to the semi-confined case, only without the channel. } 
    \label{fig:models_diagram}
\end{figure*}

The aim is to quantify the difference in outflows and energy output between a 1-D SN model and one which considers the confinement effect. To this end, we consider three models: (a) a \textit{free-field SN}, that is, a spherical blast in a uniform medium, (b) a \textit{confined SN}, whereby the blast detonates within a spherically symmetric fully-confined feedback-driven bubble, and (c) a \textit{semi-confined SN}, in which the feedback bubble is connected to the outer environment via a channel in the molecular cloud. Such comparisons provide insights on how much 1-D sub-grid models would underpredict or overpredict the actual amount of feedback. The problem in concern is depicted in Fig.~\ref{fig:models_diagram}. 

On the left, a free-field SN model is shown. The uniform medium in this scenario can represent, either, the ambient ISM, or a `smeared' molecular cloud that resembles a sub-grid model. On the right, our semi-confined model is presented. We consider a progenitor surrounded by an evolved H {\scshape ii} region, bordered by a swept-up shell of dense gas. The H {\scshape ii} region here acts as a representative of the feedback-driven cavities formed within molecular clouds. We also consider a low-density warm envelope around the molecular cloud that serves as the external medium \citep[e.g.][]{wannier83,lucas17}. This envelope represents the outer parts of the cloud that are exposed to external FUV radiation fields and are predominantly formed of atomic gas due to the lack of shielding. Its warm temperature exerts a slight thermal pressure that keeps the cloud super-virial and less likely to be completely dispersed by feedback \citep[][]{dale12b}. In reality, the transition from atomic gas to molecular in the photodissociation regions should be gradual, as the level of shielding required by $\mathrm{H_2}$ and CO differ \citep[][]{wolfire10}, but following \citet{lucas17} we assert that this simplified geometry is sufficient to serve the purpose. Crucially, there is a low-density channel that connects the cavity to this envelope. This channel is being omitted in our confined SN model.

There are a few reasons for using models with simple geometries as a first proof of concept. First, it amplifies the effect of semi-confinement, allowing easier identification. Secondly, it removes the stochastic influence from density structures developed from turbulence, allowing for a higher degree of control over the amount of venting that the SN remnant experiences. This is similar to varying the fractal dimension of the cloud \citep[e.g.][]{walch12}, but with well-defined holes that enable the rate of mass loss to be precisely calculated. Finally, it serves as a testing ground for understanding the physics behind SN energy release via channels. These findings may be then applied to turbulent cloud simulations to determine whether or not the semi-confinement effects are at play. 

\subsection{Semi-confined supernova} \label{sec:semiconf_model} 

The release of SN energy from cavities is analogous to the problem of vented explosions in engineering, which concerns detonations within partially confined environments such as rooms fitted with windows. We conjecture that their theoretical descriptions are self-similar and may be applied to astrophysics. As such, we developed our semi-confined SN model based on the work of \citet{feldgun11,feldgun16}, who studied the gas outflows and residual pressure relief from the vents of confined volumes after interior explosions. Our way to quantify the local perturbation caused by the outflow is to consider a `detector' placed at the opening of the channel (cf. Fig.~\ref{fig:models_diagram}). The following calculations model the time-evolution of gas velocity and pressure at the location of this detector. The precise values for the physical parameters involved in the equations will be given in Section~\ref{sec:setup}.

\subsubsection{Initial conditions} \label{sec:semiconf_model_setup} 

The semi-confined scenario depicted in Fig.~\ref{fig:models_diagram} sets from the time shortly after the first SN. The H {\scshape ii} region is assumed to have reached its stagnation radius $r_\mathrm{HII}$ during the end phase of D-type expansion. Suppose the current H {\scshape ii} region has volume $V_\mathrm{HII}$ and density $\rho_\mathrm{HII}$. Given that its temperature $T_\mathrm{HII}$ is typically at around $10^4\ \mathrm{K}$, we can compute the cavity’s specific internal energy $u_\mathrm{HII}$ via the relation 
\begin{equation}
    u = \frac{kT}{\mu m_H (\gamma-1)}  ,
	\label{eq:u_T_relation}
\end{equation}
where $k$ is the Boltzmann constant, $\mu$ is the mean molecular weight, $m_H$ is the hydrogen mass, and $\gamma = C_P/C_V = 5/3$ for ideal monatomic gas. From this, we obtain its gas pressure $p_\mathrm{HII}$ using the adiabatic equation of state, 
\begin{equation}
    p = \rho \left( \gamma - 1 \right) u.
    \label{eq:eos}
\end{equation}
Hence, all of the following calculations rely on an adiabatic assumption. The effects of radiative cooling are assessed here only with numerical simulations (see Section~\ref{sec:numerical_methods}); incorporating cooling timescales into the analytical model may be carried out in a future paper. We also assume that the cavity is only powered by stellar photoionization, since otherwise the cavity is likely to develop a gradient in its density profile or other shocked structures due to the presence of radiation pressure and stellar winds \citep[e.g.][]{draine11}. 

The density and radius of the molecular cloud are denoted by $\rho_\mathrm{cloud}$ and $r_\mathrm{cloud}$ respectively. Cold molecular gas is at around $10\ \mathrm{K}$, and its pressure $p_\mathrm{cloud}$ is obtained with equation~(\ref{eq:eos}). The warm envelope has density $\rho_\mathrm{env}$ and we assume its temperature to be $10^3\ \mathrm{K}$ \citep[cf.][]{lucas17}, giving its pressure $p_\mathrm{env}$. The shell of swept-up ISM gas surrounding the H {\scshape ii} region has thickness $\delta r_\mathrm{shell}$ and density $\rho_\mathrm{shell}$, hence an initial mass of $M_\mathrm{shell,0} = \rho_\mathrm{shell} 4 \pi r_\mathrm{HII}^2 \delta r_\mathrm{shell}$. Since the shell expansion has stalled at this point, $p_\mathrm{cloud}$ and $p_\mathrm{HII}$ should be approximately equal, though this is not a necessary requirement for our calculations. 

Similar to \citet{geen22}, we consider the outflow to occur only from a small solid angle from the progenitor. We model the feedback-carved channel as an opening angle from the H {\scshape ii} region. The area of the hole on the shell is hence $s_\mathrm{out} = \Omega r_\mathrm{HII}^2$, where $\Omega$ is the solid angle. However, in this analytical model, $s_\mathrm{out}$ only controls the total outflow rate from the cavity, and therefore does not distinguish between a wider solid angle and more holes on the shell. This variable is analogous to the shell cover fraction $C_f$ introduced by \citet{harperclarkemurray09}, who incorporated this parameter into the models of wind-blown bubbles \citep[][]{castor75} to account for incomplete confinement. Similarly, \citet{rahner17} also modelled the leakages through shells via an escape fraction $f_\mathrm{esc}$, though, unlike Fig.~\ref{fig:models_diagram}, they assumed that the leakage does not begin until the shell reaches the edge of the cloud, or at least a fraction of its total radius if shell fragmentation is taken into account \citep[][]{rahner18}. 

It is also worth noting that the term $s_\mathrm{out}$ in our model covers all exposed areas on the shell through which fluid can advect, regardless of its cause. It hence encompasses the existing low-density channels in the cloud and those cleared by pre-SN feedback, as well as the gaps that formed due to gravitational fragmentation \citep[e.g.][]{whitworth94b}, Rayleigh--Taylor instabilities \citep[e.g.][]{krumholz09,kumar13}, or Vishniac instability \citep[e.g.][]{miniere18}, all of which are highly probable especially during the late evolutionary stages of the bubble. The only requirement is that the area of the vent(s) must be small relative to the cavity volume. 

Consider a SN progenitor of total energy $E_\mathrm{SN} \sim 10^{51}\ \mathrm{erg}$ contained within an initially small volume $V_\mathrm{SN}$, which corresponds to the SN injection radius, located inside the cavity. We can first write down an energy conservation equation, 
\begin{equation}
    \frac{p_\mathrm{HII}}{\rho_\mathrm{HII} \left( \gamma - 1 \right) } \rho_\mathrm{HII} \left( V_\mathrm{HII} - V_\mathrm{SN} \right) + E_\mathrm{SN} = \frac{p}{\rho \left( \gamma - 1 \right) } \rho V
    \label{eq:intial_energy_conservation}
\end{equation}
\citep[cf.][]{feldgun11}. The R.H.S. gives the total energy contained within the cavity, whose density, pressure and volume are denoted by $\rho$, $p$ and $V$ respectively. The initial cavity pressure $p$ can thus be obtained by rearranging equation~(\ref{eq:intial_energy_conservation}). Its total mass $M$ is given by 
\begin{equation}
    M = \rho_\mathrm{HII} \left( V - V_\mathrm{SN} \right) + M_\mathrm{SN},
    \label{eq:intial_mass}
\end{equation}
where $M_\mathrm{SN}$ is the total ejecta mass. Combining equations~(\ref{eq:intial_energy_conservation}) and (\ref{eq:intial_mass}) hence gives an estimate of the energy per unit mass in the cavity, or the SN mass-loading,
\begin{equation}
    \epsilon = \frac{1}{M} \frac{p}{\rho \left( \gamma - 1 \right) } \rho V.
    \label{eq:mass_loading}
\end{equation}
This equation assumes that the SN energy is equally distributed amongst the gas within the cavity before the pressure relief phase begins. Following \citet{dwarkadas05}, we assume that the cavity can be simply regarded as a region of uniform internal energy. This assumption might not hold during initial stages where the shock waves are still oscillating between the shells, but by the end of this brief non-stationary phase, the SN remnant should slowly approach complete thermalization as it collides with the cavity walls \citep[see also][]{walchnaab15}. Thus, as far as pressure relief phase is concerned, the uniform cavity assumption should be appropriate. This is also in line with the `spatially lumped gas’ approximation made in \citet{feldgun11}, that the gas properties \textit{within} the confined volume varies only with time, with no radial or directional dependence.  

\subsubsection{Shell collision} \label{sec:semiconf_model_shell} 

The first part of our model considers the spherically symmetrical collision between the SN and the circumstellar shell. This problem has been first extensively studied by e.g. \citet{chevalierliang89,dwarkadas05,dwarkadas07}, who modelled the dynamics of the shell upon SN impact. Whether or not the evolution of the SN remnant would be significantly affected depends on the mass of the swept-up material relative to the ejecta, which in turn is dependent on the mass loss from the progenitor mainly during its supergiant phase. For simplicity, we adopt the thin-shell approximation, such that $\delta r_\mathrm{shell} \ll r_\mathrm{HII}$ and the shock front remains attached to the shell’s position. The following calculations are largely based on the work of \citet[Section 5]{dwarkadas05}. 

As with many self-similar solutions, the initial velocity of the shell upon collision with the SN shock can be estimated using dimensional analysis, giving
\begin{equation}
    v_\mathrm{shell,0} = \left( \frac{\beta p}{\rho_\mathrm{shell}} \right)^{1/2},
    \label{eq:shell_vel_init}
\end{equation}
scaled by a constant $\beta \sim 6$ \citep[][]{chevalierliang89}. The pressure term $p$ in equation~(\ref{eq:shell_vel_init}) originally refers to the post-shock pressure in \citet{chevalierliang89}, but here we will approximate it as the cavity pressure. To understand how the shell would evolve under the pressure imbalance between the bubble and the cloud, we write down an equation of motion, 
\begin{equation}
    F = M_\mathrm{shell}(r) \frac{dv_\mathrm{shell}}{dt} = 4 \pi r^2 p_\mathrm{net},
    \label{eq:shell_EOM}
\end{equation}
where $p_\mathrm{net}$ is the net pressure acting on each side of the shell. The variable $r$ denotes the instantaneous radius of the cavity, and initially, $r = r_\mathrm{HII}$. We can rewrite equation~(\ref{eq:shell_EOM}) as 
\begin{equation}
    \frac{dv_\mathrm{shell}}{dt} = \frac{4 \pi r^2}{M_\mathrm{shell}(r)} \left( p - \rho_\mathrm{cloud} v_\mathrm{shell}^2 \right).
    \label{eq:shell_vel_evol}
\end{equation}

Note that the mass of the shell in equations~(\ref{eq:shell_EOM}) and (\ref{eq:shell_vel_evol}) grows as a function of $r$. This is because as the cavity expands after the collision, the cold gas in the molecular cloud is continuously being swept on to the shell. We model this by, first, considering its mass conservation,
\begin{equation}
    \frac{dM_\mathrm{shell}}{dt} = 4 \pi r^2 \rho_\mathrm{cloud} v_\mathrm{shell}
    \label{eq:shell_mass_conservation}
\end{equation}
\citep[cf.][]{dwarkadas05}. Integrate by separation of variables gives 
\begin{equation}
    \int dM_\mathrm{shell} = 4 \pi \rho_\mathrm{cloud} \int r^2 \frac{dr}{dt} dt. 
    \label{eq:shell_mass_conservation_int}
\end{equation}
With this, the shell mass $M_\mathrm{shell}$ at current cavity radius $r$ can be expressed in terms of its initial value $M_\mathrm{shell,0}$,
\begin{equation}
    M_\mathrm{shell}(r) = M_\mathrm{shell,0} + 4 \pi \rho_\mathrm{cloud} \int_{r_\mathrm{HII}}^{r} r'^2 dr'
    \label{eq:shell_mass_r}
\end{equation}
The second term in the above gives the swept-up mass as cavity expands from the initial radius $r_\mathrm{HII}$ to $r$. By integrating equation~(\ref{eq:shell_mass_r}), we can arrive at the expression 
\begin{equation}
    M_\mathrm{shell}(r) = M_\mathrm{shell,0} \left[ \frac{M_\mathrm{r_{HII}}}{M_\mathrm{shell,0}} \left( \frac{r^3}{r_\mathrm{HII}} -1 \right) +1   \right],
    \label{eq:shell_totmass_final}
\end{equation}
where $M_\mathrm{r_{HII}}$ denotes the mass of surrounding medium (the molecular cloud) if it had extended from $r = 0$ to $r = r_\mathrm{HII}$, given by 
\begin{equation}
    M_\mathrm{r_{HII}} = \frac{4}{3} \pi r_\mathrm{HII}^3 \rho_\mathrm{cloud}. 
    \label{eq:mass_r0_ri}
\end{equation}

As long as the cavity is dominated by the pressure from the SN explosion, equation~(\ref{eq:shell_vel_evol}) should be sufficient to describe its motion. However, at later times the gravity acting on the shell may be able to exert an inward force and, by the end, drive an implosion that causes the shell materials to once again fill the evacuated cavity \citep[e.g.][]{romano24}. Despite the end stages of the SN remnant is beyond the scope of our model, we may roughly account for this by including an extra term to subtract off equation~(\ref{eq:shell_vel_evol}),
\begin{equation}
    \frac{dv_\mathrm{shell}}{dt} = \frac{4}{3} \pi G r \rho ,
    \label{eq:grav_on_shell}
\end{equation}
which is the gravitational acceleration towards the centre of the H {\scshape ii} region. But note that the simulations presented in this paper do not include self-gravity in order to isolate the effect of SN pressure. 

\subsubsection{Outflow rate} \label{sec:semiconf_model_outflow}

The second part of the calculations considers the venting of gas from the cavity into the outer envelope. A simple model has been proposed by \citet{feldgun11} for modelling interior explosions within containers that has a flexible cover -- we apply their approach to tackle our problem. We assume that the SN energy is only permitted to leave through a channel, and there is no energy loss via thermal conduction from the cold shell \citep[e.g.][]{keller14} or turbulent mixing at the bubble-ISM interface \citep[e.g.][]{geen23}. The only goal is to gauge the effects of limiting SN outflows to certain particular direction(s). 

Consider a streamline that leads from the inside of the cavity, just beneath the cloud's radius as marked by the red cross in Fig.~\ref{fig:models_diagram}, to an arbitrary point in the outer warm envelope. All outflows from the cavity follow this streamline, and we make the crude approximation that there is no energy loss and no development of Kelvin–Helmholtz instability along the flow. We also assume that the gas behaves adiabatically, and thus obeys Poisson’s adiabatic equation, $pV^{\gamma} = \mathrm{Const.}$, which implies that the relation
\begin{equation}
    p \propto V^{-\gamma} \propto \rho^{\gamma}
    \label{eq:poss_adiabate}
\end{equation}
would hold at any arbitrary point along the stream. The streamline flow is also governed by the Bernoulli equation,
\begin{equation}
    \frac{v'^2}{2} + \Pi \left(p'\right) = \mathrm{Const.}, 
    \label{eq:bernoulli_formula}
\end{equation}
where $v'$ is the velocity of the flow at an arbitrary point along the stream, $p'$ is its pressure, and $\Pi \left(p'\right)$ is the potential energy per unit mass at this point. The starting point of the stream located just inside the cavity therefore has pressure $p$, density $\rho$, and velocity $v_\mathrm{out}$ -- these are the properties at the detector that we aim to calculate. The end point of the stream in the envelope has pressure $p_\mathrm{env}$, density $\rho_\mathrm{env}$, and velocity $v_\mathrm{env}$. Following \citet{feldgun11}, we consider the end point to be the `reference point' in the flow, and evaluate the potential energy term $\Pi$ in equation~(\ref{eq:bernoulli_formula}) as the \textit{pressure energy} per unit mass. The potential energy at a certain point along the stream (denoted by the subscript $c$) is hence given by
\begin{equation}
    \Pi \left(p_c\right) = \int_{p_c}^{p_\mathrm{env}} \frac{1}{\rho(p')} \ dp', 
    \label{eq:pressure_energy}
\end{equation}
where the lower limit in the integral is the pressure at this particular point, $p_c$, and the upper limit is the pressure at the reference point that was defined to be $p_\mathrm{env}$. Combining equations~(\ref{eq:bernoulli_formula}) and (\ref{eq:pressure_energy}), we can now write down an equation that relates the gas properties at the two ends of the stream, 
\begin{equation}
    \frac{v_\mathrm{out}^2}{2} + \int_{p}^{p_\mathrm{env}} \frac{1}{\rho(p')} \ dp' = \frac{v_\mathrm{env}^2}{2} + \int_{p_\mathrm{env}}^{p_\mathrm{env}} \frac{1}{\rho(p')} \ dp',
    \label{eq:bernoulli_streamline}
\end{equation}
since $p_c = p$ at the starting point and $p_c = p_\mathrm{env}$ at the end point. The integral term in the R.H.S. can then be eliminated. 

The trick to solve equation~(\ref{eq:bernoulli_streamline}) is to define the end point of the streamline to where the velocity of the flow drops to zero, i.e. $v_\mathrm{env} = 0$. This further simplifies the equation to 
\begin{equation}
    \frac{v_\mathrm{out}^2}{2} + \int_{p}^{p_\mathrm{env}} \frac{1}{\rho(p')} \ dp' = 0. 
    \label{eq:bernoulli_streamline_finalform}
\end{equation}
We may now also use Poisson’s adiabatic relations (equation~\ref{eq:poss_adiabate}) to relate the density and the pressure of the cavity, which is the stream starting point, to that of the envelope, 
\begin{equation}
    p = p_\mathrm{env} \left( \frac{\rho }{\rho_\mathrm{env}} \right)^\gamma. 
    \label{eq:poss_adiabate_streamline}
\end{equation}
This equation can then be rearranged to give an expression for $\rho(p)$. Substituting it into equation~(\ref{eq:bernoulli_streamline_finalform}) and integrate by change of variables, we can arrive at the equation 
\begin{equation}
    \frac{v_\mathrm{out}^2}{2} + \frac{\gamma}{\gamma-1} \frac{p_\mathrm{env}}{\rho_\mathrm{env}} \left[ 1 - \left( \frac{p}{p_\mathrm{env}} \right)^{\frac{\gamma-1}{\gamma}} \right] = 0 
    \label{eq:streamline_vented}
\end{equation}
\citep[cf.][]{feldgun11}. Finally, rearranging equation~(\ref{eq:streamline_vented}) gives the velocity at the starting point of the stream, i.e. the outflow velocity from the channel,
\begin{equation}
    v_\mathrm{out} = \sqrt{ 2 \frac{\gamma}{\gamma-1} \frac{p_\mathrm{env}}{\rho_\mathrm{env}} \left[ \left( \frac{p}{p_\mathrm{env}} \right)^{\frac{\gamma-1}{\gamma}} - 1 \right] }. 
    \label{eq:vent_velocity}
\end{equation}
As the gas leaks through the channel, the mass loss rate from the bubble may be estimated as
\begin{equation}
    \frac{dM}{dt} = \rho_\mathrm{env} v_\mathrm{out} s_\mathrm{out}
    \label{eq:mass_loss}
\end{equation}
where $s_\mathrm{out}$ is the total area of the hole(s) on the shell. 

If the envelope properties remain constant, $v_\mathrm{out}$ (equation~\ref{eq:vent_velocity}) varies only with the instantaneous cavity pressure $p$. As mentioned earlier, it does not require whether $s_\mathrm{out}$ is formed of a single hole or multiple smaller holes. For the latter case, this model predicts the same $v_\mathrm{out}$ in outflows from all vents, though, of course, the mass loss rate (equation~\ref{eq:mass_loss}) would be higher. However, these vents must be sufficiently small for the outflows to be considered as a single streamline, which otherwise the use of Bernoulli equation would be invalid. As such, our model cannot account for massive bulk outflows in the case of very large openings. Furthermore, there may be non-linear effects when outflows through different channels interact with each other. Since the above calculations are independent of the orientation of the vents, this scenario may only be tested using 3-D numerical simulations. 

\subsubsection{Adiabatic pressure relief} \label{sec:semiconf_model_relief} 

The third part considers the rate of pressure decay in the cavity as the SN energy leaves. We make a crude assumption that the energy per unit mass within the cavity, as initially deposited from the SN, remains the same throughout the evolution. We also assume that the Poisson’s adiabatic relation (equation~\ref{eq:poss_adiabate_streamline}) governing the streamline continues to hold during this pressure relief phase. As the cavity radius increases (equation~\ref{eq:shell_vel_evol}) and the mass drops (equation~\ref{eq:mass_loss}), we update the cavity pressure using its current mass $M$ and radius $r$, 
\begin{equation}
    p = \epsilon \left( \gamma-1 \right) \frac{M}{V}
    \label{eq:cavity_pressure_update}
\end{equation}
where $V = 4/3 \pi r^3$ and $\epsilon$ is from equation~(\ref{eq:mass_loading}). Substituting this into the Poisson's adiabatic relation gives the new cavity density, 
\begin{equation}
    \rho = \rho_\mathrm{env} \left( \frac{p}{p_\mathrm{env}} \right)^{1/\gamma}.
    \label{eq:cavity_density_update}
\end{equation}

Finally, we put all equations described in Section~\ref{sec:semiconf_model_setup}, \ref{sec:semiconf_model_shell}, \ref{sec:semiconf_model_outflow} and \ref{sec:semiconf_model_relief} together and evolve this system of equations numerically to give the time-evolution of gas properties at the detector. The steps to initialize the parameters and the way to update them in each time-step are summarized in Table~\ref{tab:analytical_model_init} and \ref{tab:analytical_model_evol}. Here, we adopt a simple Eulerian integrator to evolve the parameters. Using velocity Verlet or Runge-Kutta integrators would be more ideal, but no numerical stability issues have been encountered so far. The calculation outputs for $\gamma=5/3$ are presented in Section~\ref{sec:outflow} to compare against the results from 3-D hydro simulations\footnote{However, considering that the SN is colliding with molecular gas that surrounds the cavity, we could alternatively set $\gamma = 7/5$ (for diatomic gas) to gauge the behaviour of this gas mixture. It can be shown with our analytical model that using $\gamma = 7/5$ would slightly decrease the velocity of the outflows at the detector location, but otherwise the effect is minimal.}. We will show that this analytical model is capable of predicting the outflow properties in spite of the approximations and simplifications made in the calculations. 

\begin{table}
    \centering
    \caption{Algorithm to initialize the analytical model of a semi-confined SN}
    \label{tab:analytical_model_init}
    \begin{tabular}{lccr} 
        \hline
        Procedure & Equation \\
        \hline
        Calculate initial pressure $p$ & \ref{eq:intial_energy_conservation} \\
        Calculate initial mass within cavity $M$ & \ref{eq:intial_mass} \\
        Calculate mass loading $\epsilon$ & \ref{eq:mass_loading} \\ 
        Initiate $r$ by setting it to $r_\mathrm{HII}$ \\
        Calculate initial shell velocity $v_\mathrm{shell}$ & \ref{eq:shell_vel_init} \\
        \hline
    \end{tabular}
\end{table}

\begin{table}
    \centering
    \caption{Algorithm to evolve the analytical model of a semi-confined SN}
    \label{tab:analytical_model_evol}
    \begin{tabular}{lccr} 
        \hline
        Procedure in each time-step $dt$ & Equation \\
        \hline
            Calculate $v_\mathrm{out}(p)$ & \ref{eq:vent_velocity} \\
            Calculate $dM/dt$ & \ref{eq:mass_loss} \\
            Update $M = M - \frac{dM}{dt} dt$  \\
            Calculate mass of shell + mass swept $M_\mathrm{shell}$ & \ref{eq:shell_totmass_final} \\
            Calculate acceleration of shell $ dv_\mathrm{shell} / dt$ & \ref{eq:shell_vel_evol} \\
            Update velocity $v_\mathrm{shell} = v_\mathrm{shell} + \left( dv_\mathrm{shell}/dt \right) dt$ \\ 
            Update cavity radius $r = r + v_\mathrm{shell} dt $ \\
            Calculate cavity pressure $p$ & \ref{eq:cavity_pressure_update} \\
            Calculate cavity density $\rho$ & \ref{eq:cavity_density_update} \\
        \hline
    \end{tabular}
\end{table}

\subsection{Free-field supernova} \label{sec:freefield_model} 

We now turn to model the outflows from a free-field SN, where there is no preferential energy escape route. The explosion occurs within a uniform ambient medium of density $\rho_\mathrm{amb}$. To allow for comparisons with the semi-confined model, we consider a detector that is placed at the same distance away from the progenitor, as indicated in Fig.~\ref{fig:models_diagram}. We model the evolution of gas properties at the detector \textit{after} the SN shock front has passed through it, meaning the detector is located in the post-shock region. Since the blast is now spherically symmetrical, the well-established self-similar shock wave equations apply. The Sedov--Taylor solution \citep[][]{sedov46,taylor50} gives the velocity of the shock front $v_\mathrm{shock}$ as a function of its radius $r_\mathrm{shock}$,  
\begin{equation}
    v_\mathrm{shock} = \frac{2}{5} \xi_0^{5/2} \left( \frac{E_\mathrm{SN}}{\rho_\mathrm{amb}} \right)^{1/2} r_\mathrm{shock}^{-3/2}
    \label{eq:sedov_velocity}
\end{equation}
where $\xi_0$ is a dimensionless variable. For $\gamma = 5/3$, we can assume $\xi_0 \sim 1.15$ \citep[][]{ostrikermckee88}. 

To determine the properties behind this shock front, we first define a scale-free radius $\eta$, 
\begin{equation}
    \eta = \frac{r_0}{r_\mathrm{shock}} 
    \label{eq:scale_free_rad}
\end{equation}
\citep[cf.][]{diazrigby22}, where $r_0$ denotes the position of the detector ($r_0 < r_\mathrm{shock}$). The gas velocity, density and pressure at the detector are denoted by $v_0$, $\rho_0$ and $p_0$ respectively. These properties can be calculated by applying the Rankine--Hugoniot relations \citep[][]{rankine1870,hugoniot87}. These relations are formed based on boundary conditions on the hydro equations at the shock front discontinuity, derived from the conservation of mass, energy and momentum. In terms of variables considered in our model, the solutions read:
\begin{equation}
    \frac{v_0}{v_\mathrm{shock}} = \frac{\eta}{\gamma} + \left( \frac{\gamma - 1}{\gamma^2 + \gamma} \right) \eta^a ,
    \label{eq:vel_behind_shock}
\end{equation}
\begin{equation}
    \frac{\rho_0}{\rho_\mathrm{amb}} = \left( \frac{\gamma+1}{\gamma-1}  \right) \frac{\eta^b}{\gamma^c} \left( \gamma + 1 - \eta^{a-1} \right)^c ,
    \label{eq:rho_behind_shock}
\end{equation}
\begin{equation}
    \frac{p_0}{p_\mathrm{amb}} = M_s^2 \left( \frac{2\gamma^{1-d}}{\gamma+1} \right) \left( \gamma + 1 - \eta^{a-1} \right)^d ,
    \label{eq:pressure_behind_shock}
\end{equation}
with the exponents given by
\begin{equation}
    a = \frac{7\gamma-1}{\gamma^2-1}, \\
    b = \frac{3}{\gamma-1}, \\
    c = \frac{2\gamma+10}{\gamma-7}, \\
    d = \frac{2\gamma^2+7\gamma-3}{\gamma-7} \\
    \label{eq:shock_eq_exponents}
\end{equation}
\citep[cf.][equation 16-19]{diazrigby22}. The Mach number $M_s$ in equation~(\ref{eq:pressure_behind_shock}) is given by the ratio of shock velocity to the sound speed,
\begin{equation}
    M_s = \frac{v_\mathrm{shock}}{c_s},
    \label{eq:mach_number}
\end{equation}
where $c_s = \sqrt{\gamma p_\mathrm{amb}/\rho_\mathrm{amb}}$. We assume that the above relations hold throughout the Sedov phase, and we do not consider its transition to the snowplough phases to be consistent with the adiabatic assumptions used in our semi-confined model. 

Similar to before, we evolve the equations numerically to solve for the time-evolution of the gas properties. The shock radius $r_\mathrm{shock}$ begins at the position of the detector $r_0$. The subsequent steps to evolve the system are summarized in Table~\ref{tab:analytical_model_evol_ff}. The calculation results are presented in Section~\ref{sec:outflow} to compare against the semi-confined model and the results from 3-D numerical simulations. 

\begin{table}
    \centering
    \caption{Algorithm to evolve the analytical model of a free-field SN}
    \label{tab:analytical_model_evol_ff}
    \begin{tabular}{lccr} 
        \hline
        Procedure in each time-step $dt$ & Equation \\
        \hline
            Calculate $v_\mathrm{shock}$ & \ref{eq:sedov_velocity} \\
            Update $r_\mathrm{shock} = r_\mathrm{shock} + v_\mathrm{shock} dt$ \\ 
            Calculate $\eta$ & \ref{eq:scale_free_rad} \\
            Calculate $M_s$ & \ref{eq:mach_number} \\ 
            Calculate $v_0$, $\rho_0$ and $p_0$ & \ref{eq:vel_behind_shock},\ref{eq:rho_behind_shock},\ref{eq:pressure_behind_shock} \\ 
        \hline
    \end{tabular}
\end{table}

\subsection{Confined supernova} \label{sec:confined_model} 

A fully-confined SN may be modelled in a similar way to the free-field case. We assume that once the SN shock collides with the shell of the H {\scshape ii} region, it immediately transitions to its snowplough phase \citep[e.g.][]{walchnaab15} and the shock front continues to expand into the cloud. Our detector is placed beyond the H {\scshape ii} region but just within the molecular cloud's boundaries. 

The snowplough phase here is assumed to be purely driven by momentum-conservation, since collisions thermalize the kinetic energy and radiative cooling should rapidly dominate once the shock reaches the dense cloud\footnote{Even in the adiabatic runs, the energy should quickly dissipate via shock heating. The dissipated energy may not be concentrated in the remnant interior to continue to drive the shock with its thermal pressure.}. To model the remnant's behaviour at this stage, we first consider the momentum $\mu$ carried by the shell immediately before it encounters the cavity walls. This can be determined by simply multiplying $M_\mathrm{shell,0}$ (defined in section~\ref{sec:semiconf_model_setup}) by the Sedov velocity (equation~\ref{eq:sedov_velocity}) at $r_\mathrm{shock} = r_\mathrm{HII}$, with $\rho_\mathrm{amb} = \rho_\mathrm{HII}$. Momentum $\mu$ has dimension $[MLT^{-1}]$. We can hence write the characteristic length of the shock (impact radius) as a function of time, 
\begin{equation}
    r_\mathrm{shock} = \beta_0 \left( \frac{\mu t}{\rho_\mathrm{cloud}} \right)^{1/4}
    \label{eq:snowplough_radius}
\end{equation}
\citep[cf. e.g.][]{oort51}, where $\beta_0$ is another dimensionless variable of order unity, analogous to $\xi_0$ in equation~(\ref{eq:sedov_velocity}); we assume $\beta_0 \sim 1$. Velocity of the shock front is simply the time-derivative of $r_\mathrm{shock}$, giving
\begin{equation}
    \begin{aligned}
    v_\mathrm{shock} & = \frac{1}{4} \beta_0 \left( \frac{\mu}{\rho_\mathrm{cloud}} \right)^{1/4} t^{-3/4} \\
    & = \frac{1}{4} \beta_0^{-2} \left( \frac{\mu}{\rho_\mathrm{cloud}} \right)  r_\mathrm{shock}^{-3}.
    \end{aligned}
    \label{eq:snowplough_velocity}
\end{equation} 

Since equations~(\ref{eq:snowplough_radius}) and (\ref{eq:snowplough_velocity}) are self-similar, we can apply the Rankine--Hugoniot relations (equations~\ref{eq:vel_behind_shock}-\ref{eq:shock_eq_exponents}) to compute the physical properties in its post-shock region. The only difference is that, here, the shock is propagating from within the cloud, hence $\rho_\mathrm{amb} = \rho_\mathrm{cloud}$ and $p_\mathrm{amb} = p_\mathrm{cloud}$. We neglect the transition to the outer envelope. Indeed, our simulations show that the remnant never reaches the envelope if it is fully confined.

\section{Numerical methods} \label{sec:numerical_methods} 

We perform 3-D Smoothed Particle Hydrodynamics (SPH) simulations \citep[][]{gingoldmonaghan77,lucy77} to test our analytical models as well as to further examine the differences between free-field SNe, confined SNe and semi-confined SNe. SPH is a Lagrangian particle-based method that solves the equations of hydrodynamics by discretizing the fluid into individual particles, each serving as an interpolation (sampling) point of the underlying fluid properties. The simulations are carried out with the SPH code {\scshape phantom}, developed by \citet{phantom18}, which models compressible flows and adopts a variable smoothing length formalism with equal mass particles. Shock discontinuities are treated with artificial viscosity and artificial thermal conductivity \citep[][]{chowmonaghan97,price08}. We neglect the self-gravity of the particles. 

\subsection{Supernova model}

Supernova is injected by adding extra particles whose total mass corresponds to the ejecta mass, which we roughly assume to be a quarter of that of the progenitor. The extra particles carry a total energy of approximately $10^{51}\ \mathrm{erg}$, and are randomly (but symmetrically across the Cartesian axes) positioned within a sphere of $r_\mathrm{SN}$, which we define to be $0.1\ \mathrm{pc}$. The split of thermal to kinetic energy may be varied in our code. In this paper, we inject SNe as pure kinetic energy to comply with the findings of \citet{fichtner24}, who showed that, within $5\ \mathrm{pc}$, the majority of the SNe energy should still be in kinetic form. The ejecta particles' radial velocities are set in the form of a skewed Gaussian function\footnote{We borrow the functional form of the $1^\mathrm{st}$ derivative of the SPH cubic B-spline kernel \citep[][]{schoenberg46} to set the radial velocity profile. The code scales the amplitude during runtime such that the total kinetic energy carried by the SN ejecta particles matches the required value. }. The purpose is to allow the particles closer to the centre to catch up with those further away and thus forming a shell, while also leaving some particles behind to avoid completely evacuating the centre, which can lead to poor resolution in SPH. 

Radiative cooling is incorporated using the cooling curve of \citet[Fig. 1]{joungmaclow06}. Cooling curves are pre-computed cooling rates $\Lambda$ as a function of gas temperature $T$ that replace on-the-fly chemistry calculations. This method is largely appropriate as far as gas dynamics is concerned. The \citet{joungmaclow06} curve assumes ionization equilibrium and is applicable to optically thin plasmas with solar metallicities. The high temperature regime of the curve ($T \ge 2\times10^4\ \mathrm{K}$) is from \citet{sutherlanddopita93} and the low temperature regime ($T < 2\times10^4\ \mathrm{K}$) is from \citet{dalgarnomccray72} for an ionization fraction of $10^{-2}$. 

Following \citet{koyamainutsuka02}, we also apply a small constant heating term, $\Gamma = 2 \times 10^{-26}\ \mathrm{erg\ s^{-1}}$, which covers the heating from external cosmic rays and FUV fields \citep[][]{wolfire95}. The internal energy $u$ of each particle is subsequently updated using an implicit method introduced by \citet{vazquezsemadeni07}, 
\begin{equation} 
    u_{i+1} = u_\mathrm{eq} + (u_{i} - u_\mathrm{eq}) \ \exp \left( - \frac{\mathrm{d}t}{\tau} \right),
    \label{eq:implicit_new_u}
\end{equation}
where $i$ denotes an arbitrary hydro step and $\mathrm{d}t$ is the time-step. $u_\mathrm{eq}$ is the internal energy that corresponds to the particle's equilibrium temperature. The $u_\mathrm{eq}$ values are pre-computed by solving the thermal equilibrium, $n_H\Gamma - n_H^2\Lambda(T) = 0$, where  $n_H$ is the hydrogen number density\footnote{This property relates to the particle density $\rho$ by $n_H = \rho/m_H$, where $m_H$ is the hydrogen mass.}. The equilibrium temperatures as a function of density $\rho$ are plotted in Appendix~\ref{appen:equil_temp}. $\tau$ is the timescale over which the gas particle radiates its thermal excess, given by
\begin{equation} 
    \tau = \left| \frac{u-u_\mathrm{eq}}{n_H\Gamma - n_H^2\Lambda} \right|
    \label{eq:heatcool_timescale}
\end{equation}
\citep[cf.][]{vazquezsemadeni07}. The use of implicit methods directly corrects the particle internal energy by the end of each hydro step and prevents the time-step from being overly constrained as the SN shock cools. This cooling method has also been used in \citet{bonnell13} to account for the bi-stable nature of ISM temperatures.

\subsection{Photoionization}

We incorporate stellar photoionization by employing the hybrid radiation hydrodynamics (RHD) scheme developed by \citet{petkova21,lau25}. This RHD scheme couples the SPH code {\scshape phantom} to the grid-based Monte Carlo Radiative Transport (MCRT) code {\scshape cmacionize} \citep[][]{vandenbrouckewood18,vandenbrouckecamps20} to model the thermodynamical impact of ionizing radiation in SPH. It works by transferring the particle-interpolated densities on to a Voronoi grid \citep[][]{voronoi1908} at each hydro step with the Exact density mapping method \citep[][]{petkova18}, and executing the MCRT simulation on this Voronoi density field to compute the ionic fraction of each cell. The results are subsequently returned to {\scshape phantom} for computing the photoionization heating and cooling rates for each particle \citep[cf.][]{lau25}. This scheme also employs a tree-based algorithm that adaptively converts gravity tree nodes into pseudo-SPH particles, with which we carry out the mapping, allowing resolutions to be lowered in the neutral regions. We use this method to self-consistently generate expanding H {\scshape ii} regions in our simulated molecular clouds that sweep its surrounding gas into thin but dense shells, replicating the scenario shown in Fig.~\ref{fig:models_diagram}.

\section{Results}  \label{sec:results} 

\begin{figure*}
    \includegraphics[width=6.7in]{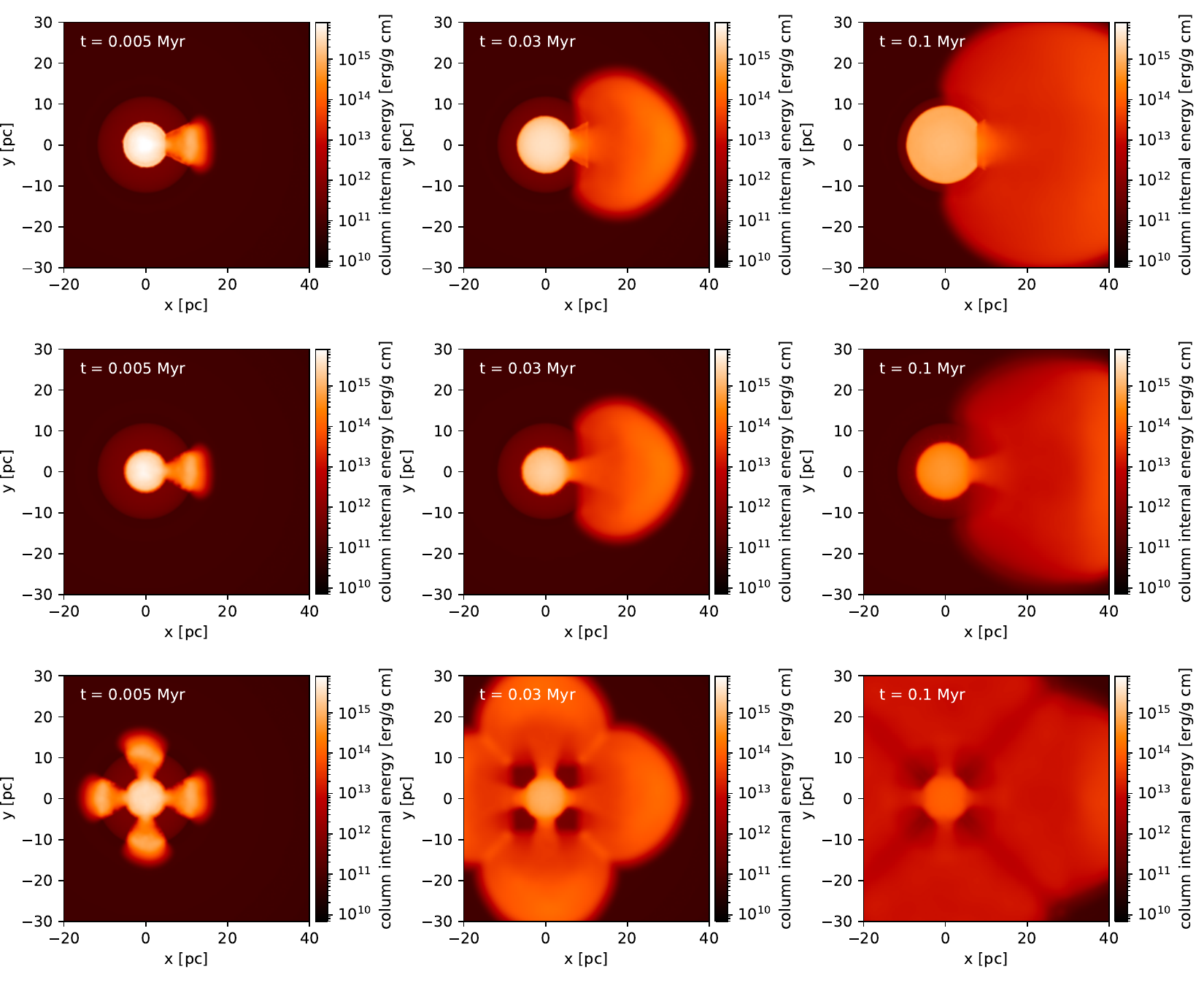}
    \caption{\textit{Left to right}: Evolution of a semi-confined SN from 3-D hydrodynamical simulations. \textit{Top}: Adiabatic run with one channel. \textit{Middle}: Cooling run with one channel. \textit{Bottom}: Cooling run with four channels.} The render shows the density-weighted column specific internal energy interpolated from the SPH particles. The high-energy SN remnant within the cavity is mostly in white, surrounded by a spherical molecular cloud that is seen as a ring in dark red. The warm envelope is seen as the faint background.  
    \label{fig:sncavity_evol}
\end{figure*}

Two sets of simulations are presented – one as an adiabatic test, and one with radiative cooling incorporated. The early stages of SN evolution may be approximated as an adiabatic process, but radiative cooling soon begins to dominate after the remnant reaches its snowplough phases. We anticipate the adiabatic results to match the predictions derived from the analytical models, whereas the results from the cooling runs would provide useful information on their realistic impact. Parsec-scale simulations usually do not suffer from numerical overcooling, nevertheless it may be assumed that the `true' effect would lie somewhere in between the two test cases. Fig.~\ref{fig:sncavity_evol} illustrates the evolution of a semi-confined SN in the adiabatic run and in the cooling run. We also show the cooling run for a semi-confined case with four identical channels. The particle density-weighted column specific internal energies are plotted, and the geometrical features seen in the images correspond to the depiction in Fig.~\ref{fig:models_diagram}.  

\subsection{Simulation initial conditions} \label{sec:setup}

The initial conditions for the adiabatic runs and the cooling runs are identical. We describe below the details of the simulation setup and the starting conditions. The detector, as illustrated in Fig.~\ref{fig:models_diagram}, is placed at $10\ \mathrm{pc}$ away from the progenitor along the x-axis.

\subsubsection{Confined model} \label{sec:conf_setup}

The confined SN setup consists of a spherical molecular cloud with radius $r_\mathrm{cloud} = 12\ \mathrm{pc}$ and density $\rho_\mathrm{cloud} = 10^{-21}\ \mathrm{g\ cm^{-3}}$ at temperature $T_\mathrm{cloud} \simeq 10\ \mathrm{K}$. Its total mass is approximately $10^5\ \mathrm{M_\odot}$, with a mass resolution of $10^{-2}\ \mathrm{M_\odot}$. No turbulence is added. A $50\ \mathrm{M_\odot}$ sink particle is placed at the centre of the cloud to be the ionizing source and the SN progenitor (hence ejecta mass $m_\mathrm{SN} = 12.5\ \mathrm{M_\odot}$). We assume an ionizing flux of $Q = 10^{51}\ \mathrm{s^{-1}}$, though this parameter only affects the initial size of the H {\scshape ii} region. The outer warm envelope has density $\rho_\mathrm{env} = 4\times10^{-25}\ \mathrm{g\ cm^{-3}}$ and a temperature of $10^3\ \mathrm{K}$.  

The H {\scshape ii} region is evolved until the ionization front reaches approximately $r_\mathrm{HII} = 4.8\ \mathrm{pc}$. This radius is smaller than the typical cavity size \citep[cf. e.g.][]{dale14,fichtner24} but could represent that of an earlier epoch, say around $6\ \mathrm{Myr}$ from the onset of feedback \citep[cf.][Fig. 1]{dale14}, which is well sufficient for the most massive stars to explode as SN. The ionized gas temperature is at around $10^4\ \mathrm{K}$ once in thermal equilibrium. The thickness of the shell was found to be around $\delta r_\mathrm{shell} \simeq 0.2\ \mathrm{pc}$ with density $\rho_\mathrm{shell} \simeq 10^{-19}\ \mathrm{g\ cm^{-3}}$.  

We further manually reduce the density of the evolved H {\scshape ii} region in order to suppress the cooling within the cavity. This is done by replacing the particles within the shell's inner boundaries with a new set of uniformly-distributed particles at temperature $10^4\ \mathrm{K}$ and density $\rho = 10^{-23}\ \mathrm{g\ cm^{-3}}$. Applying this procedure may lead to some inconsistencies in the gas evolution, but we argue that a lower density cavity is justified since, in reality, the bubbles are also subjected to stellar winds that drive materials out of the cavity. It does not affect the credibility of this simulation for testing our analytical models, given that the pressure difference between the cavity and the envelope is the only determining factor.

\subsubsection{Semi-confined model} \label{sec:semiconf_setup}

The semi-confined model is identical to that of the confined model, except it has one or more channels manually carved in the cloud. This is done by randomly deleting 99\% of the particles that lie within the cone drawn by a small solid angle, $\Omega = 0.05\times4\pi\ \mathrm{sr}$, from the location of the progenitor up to the cloud's surface. A caveat to note is that doing this effectively reduces the fluid resolution along the channel(s), which might lead to ambiguous results regarding the gas flows around these holes. This issue has also been noted in \citet{dalebonnell08} and is likely a common issue with modelling voided regions in SPH.

\subsubsection{Free-field model} \label{eq:freefield_setup}

The density of the ambient medium here is set to be the same as that of the warm envelope in the confined and semi-confined cases, so $\rho_\mathrm{amb} = 4\times10^{-25}\ \mathrm{g\ cm^{-3}}$. A higher density could have been used if the medium represents a smeared molecular cloud, however, we conjecture that a denser medium would only further restrict the propagation of the shock which results in weaker outflows. Hence, testing a free-field SN in a low-density medium provides the upper limit in the strength of its perturbation in a particular direction.

\subsection{Shell expansion} \label{sec:shellevol}

We first examine whether or not the predicted evolution of cavity radius $r$ in our semi-confined model (Section~\ref{sec:semiconf_model_shell}) matches that from the simulations. An issue is that the boundaries of the shell or shock front in the simulation may become ambiguous due to the multiple shock waves produced from the collisions between the SN and the cold molecular cloud gas. This is particularly severe in the adiabatic runs, in which the shock front extends significantly beyond the initial boundaries of the cloud itself. Furthermore, their outflows from the channel were seen to induce reverse shocks that wrapped around the other regions of the cloud and interfered with the expanding cavity. All of these factors complicate the geometry of the shell in our simulations. 

\begin{figure}
    \includegraphics[width=3.3in]{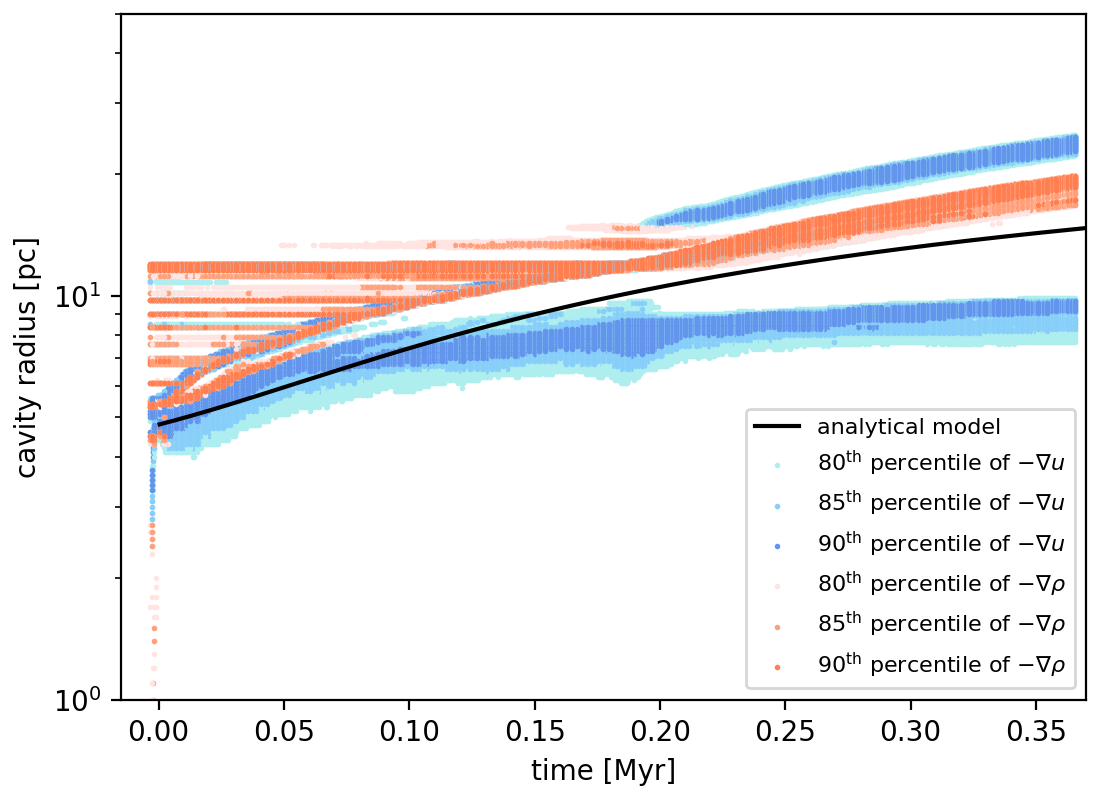}
    \caption{Expansion of cavity shell radius upon collision with the SN shock in the adiabatic semi-confined model. Prediction from the analytical model is plotted in black. The shell location in the simulation is traced by large negative gradients in density (red) or in internal energy (blue) along a 1-D line from the progenitor. Colours indicate the percentile of the gradient, with darker colours indicate steeper falls. The cloud has radius $12\ \mathrm{pc}$, and its boundary is seen as the horizontal features in the density gradients. } 
    \label{fig:shell_evol_adiabatic}
\end{figure}

\begin{figure}
    \includegraphics[width=3.3in]{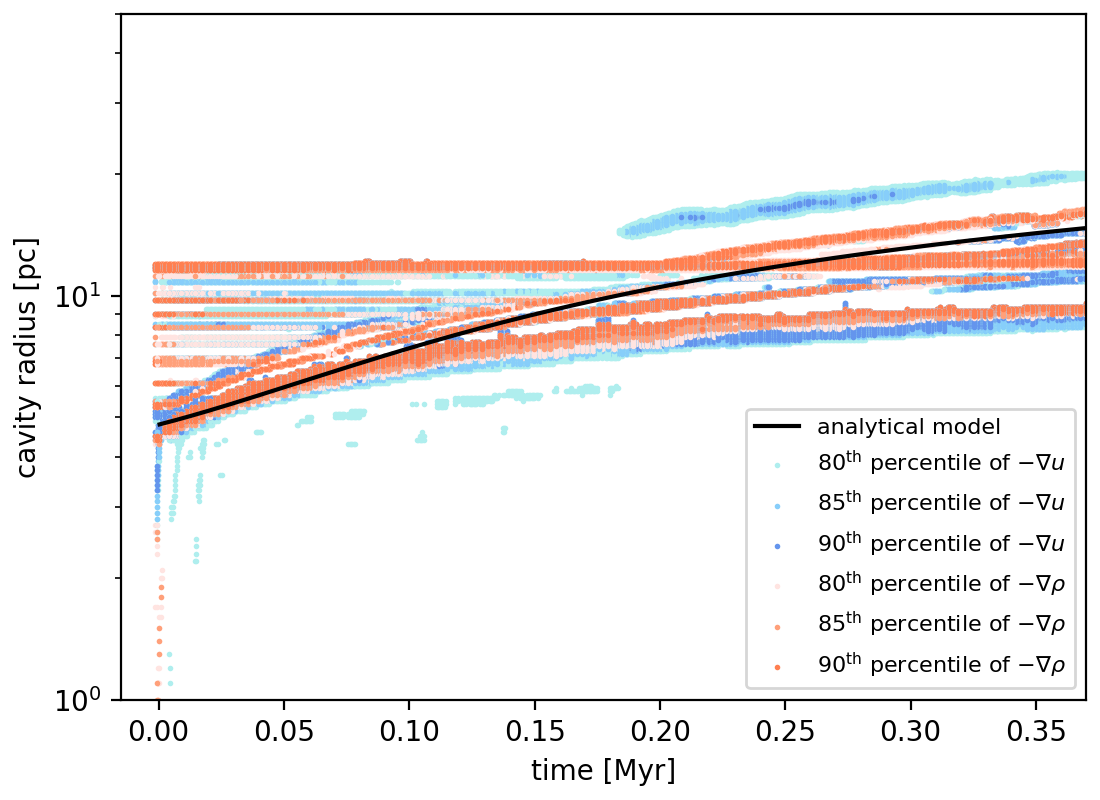}
    \caption{Same as Fig.~\ref{fig:shell_evol_adiabatic} but for the cooling run. }
    \label{fig:shell_evol_cooling}
\end{figure}

As such, we resolve to estimate the location of the shock front(s) by locating regions where large gradients in density and/or internal energy are present. We do so by, first, defining a straight line along the negative x-axis, radially away from the progenitor and opposite to the channel. Equally-spaced interpolation points are set along the line, and we measure the density and internal energy of the fluid via interpolation from neighbouring particles. From this, we obtain the density and internal energy gradients along the line, and extract the radii with the steepest gradients. The results are shown Fig.~\ref{fig:shell_evol_adiabatic} and \ref{fig:shell_evol_cooling}. 

Fig.~\ref{fig:shell_evol_adiabatic} plots the radii in the adiabatic run whose negative gradients in density and internal energy lie in their $80^\mathrm{th}-$, $85^\mathrm{th}-$ and $90^\mathrm{th}-$percentile, giving a 1-D contour map of the shock at each time-step. It highlights one side of the compressed wave. We choose to display the negative gradients rather than the positive since the negative component appears to be a better tracer of the shell boundary, whereas the positives are heavily contaminated by the energy distribution of the SN remnant itself. Note that the horizontal features seen at the early times in density gradients are the transition regions from the cloud to the envelope. 

We plot on top of this contour the results computed with our analytical model. The shape of the curve agrees very well with the shell domain traced by the density gradients. The large gap seen in the internal energy negative gradients indicates that internal energy tends to rise sharply and builds up at the inner cavity walls. The fact that the curve lies mostly along this gap arguably serves as an extra proof of our analytical model. 

We repeat the same procedures for the cooling run, as shown in Fig.~\ref{fig:shell_evol_cooling}. In this plot, the curve appears to be over-predicting the radius of the shell, assuming that it is predominantly traced by negative density gradients. This result is not surprising, since radiative losses could significantly weaken the strength of the shocks and thus reduce its ability to push against the cavity walls. The gap in internal energy is also much less prominent compared to the adiabatic run, which likely indicates that the SN energy becomes more uniformly distributed within the cavity when the shocks are allowed to cool. 

One aspect to note from Fig.~\ref{fig:shell_evol_adiabatic} is that, immediately after the SN explosion, the cavity radius undergoes a rapid expansion before transitioning into a steady rise. Despite our model successfully predicts the overall trend, this behaviour was not being reproduced. The reason likely lies in our assumption that the SN energy has already reached a quasi-steady state by the time the gas venting begins, and the direct pressure exerted by the SN was not being taken into account. It is for the same reason that this rapid expansion phase is not seen in Fig.~\ref{fig:shell_evol_cooling}, since radiative cooling has reduced the ram pressure.

\subsection{Outflows} \label{sec:outflow}

We turn to look at the evolution of gas properties at the detector in the outflow, as illustrated in Fig.~\ref{fig:models_diagram}. It provides information on the \textit{local} perturbation in the post-shock regions. If one thinks of a sub-grid model as a black box which encapsulates the volume within the cloud's boundaries, then these results would indicate the disturbance in the gas within the box after the energy has escaped to the outer ISM. They also show the strength of the outflow at the box boundary. 

Here, we compare three different physical properties: the velocity $v$ (equation~(\ref{eq:vent_velocity}) from semi-confined model; equation~(\ref{eq:snowplough_velocity}) from confined model; equation~(\ref{eq:vel_behind_shock}) from free-field model), the thermal pressure $p$ defined by equation~(\ref{eq:eos}) (equation~(\ref{eq:cavity_pressure_update}) from semi-confined model; equation~(\ref{eq:pressure_behind_shock}) from confined and free-field model), and ram pressure $p_\mathrm{ram} \approx \rho\ v^2$. These properties are obtained via an interpolation from neighbouring particles around the detector. Note also that, in our analytical model, the `spatially lumped' assumption asserts that the cavity pressure $p$ remains the same throughout the channel -- from the cavity interior to where it interfaces the warm envelope. As such, the detector placed in the simulation can be located anywhere along the channel. In our tests, we set it to roughly halfway along the tube. The detector location does not move to follow the expansion of the cavity, but remains within the spatial domains of the channel. 

\begin{figure}
    \includegraphics[width=3.2in]{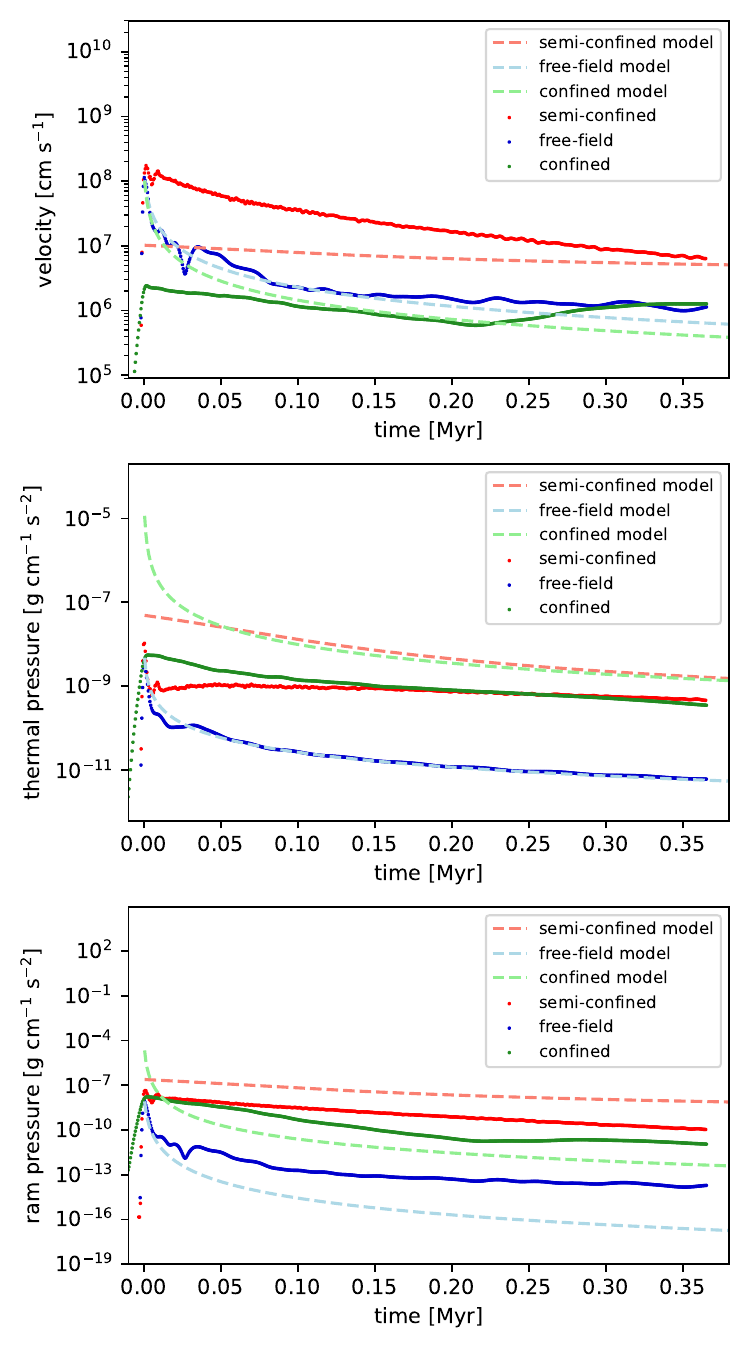}
    \caption{Gas velocity, thermal pressure and ram pressure measured at the detector location from the adiabatic simulation, plotted for the semi-confined case (red), the confined case (green), and the free-field case (blue). The curves are synchronized at the time when the shock front arrives at the detector. The results from the analytical models are shown in dotted lines for comparison. }
    \label{fig:outflow_adiabatic}
\end{figure}

Fig.~\ref{fig:outflow_adiabatic} plots the evolution for the adiabatic runs, synchronized at the time when the shock front arrives at the detector. For each property, we also show the results from the analytical model, computed by substituting the parameter values stated in Section~\ref{sec:setup} into the procedures from Tables~\ref{tab:analytical_model_init}, \ref{tab:analytical_model_evol} and \ref{tab:analytical_model_evol_ff}. We see from Fig.~\ref{fig:outflow_adiabatic} first panel that the gas velocities measured at the detector is close to their predicted values for the semi-confined case, especially at the later times. This result validates our analytical model, and demonstrates that outflows from an opening on the bubble can indeed be treated as a single streamline. The gas velocities are initially much higher than predicted due to the direct impact from the SN shock, which was not being considered in our calculations. The decay rate also appears to be slightly higher than our analytical model, and we attribute this to the expansion of the channel's cross-sectional area as the energy leaves the cavity (cf. Fig.~\ref{fig:sncavity_evol}). Section~\ref{sec:outflow_chnlsize} discusses further on this. Formulating a method to describe its evolution as a function of $v_\mathrm{out}$ would be ideal. 

As for the free-field, strong agreement has been achieved with the blast wave solutions, as seen in the first panel of Fig.~\ref{fig:outflow_adiabatic}. The oscillations likely originate from the weaker reflected and transmitted waves within the post-shock region which in turn collide with the shock front \citep[e.g.][]{cioffi88,thornton98}. More importantly, their velocities at the detector are consistently lower than the semi-confined case. The velocities in the confined case are even lower. In fact, no outflows have been generated, but the cavity expanded slightly beyond the cloud's boundaries which gave rise to the measurements at the detector location. It also shows that 1-D simulations which account for the confinement by cavities but neglect the channels can significantly underestimate the dynamical impact of SNe.  

Keeping in mind that the velocities here are that of the post-shock region, a lower velocity does not necessarily imply a slower shock front or a weaker outflow. However, it \textit{does} imply that the local ISM is being relatively less perturbed after the shock front has swept through the area. From this panel, we see that the outflows from the semi-confined case creates a higher velocity disturbance at smaller length-scales compared to the free-field. Their slower rate of decay also imply a more extended momentum deposition timescale (see Section~\ref{sec:energy_momentum}). 

The second panel in Fig.~\ref{fig:outflow_adiabatic} models the pressure relief from the (semi-)confined cavity. Their rates of decay in the simulated results agree well with the analytical models, with only a systematic offset which might have originated from overestimated densities. It also means that the thermal pressure retained at regions near the progenitor is much higher than if the remnant is freely expanding, indicating a greater local energy deposition. 

\citet{harperclarkemurray09} and \citet{lopez11} proposed that physical leakages are the primary cause of the X-ray gas pressure loss in the feedback-driven bubbles. Indeed, this is likely the case compared to a fully confined bubble. However, our results here suggest that as long as the bubble is predominantly confined, the rate of pressure loss is not significant. Likewise, ram pressure is also being sustained at a higher level, as shown in the third panel. Overall, these results demonstrate that partial confinement effects are crucial in determining the local gas pressure around SN remnants. 

\begin{figure}
    \includegraphics[width=3.2in]{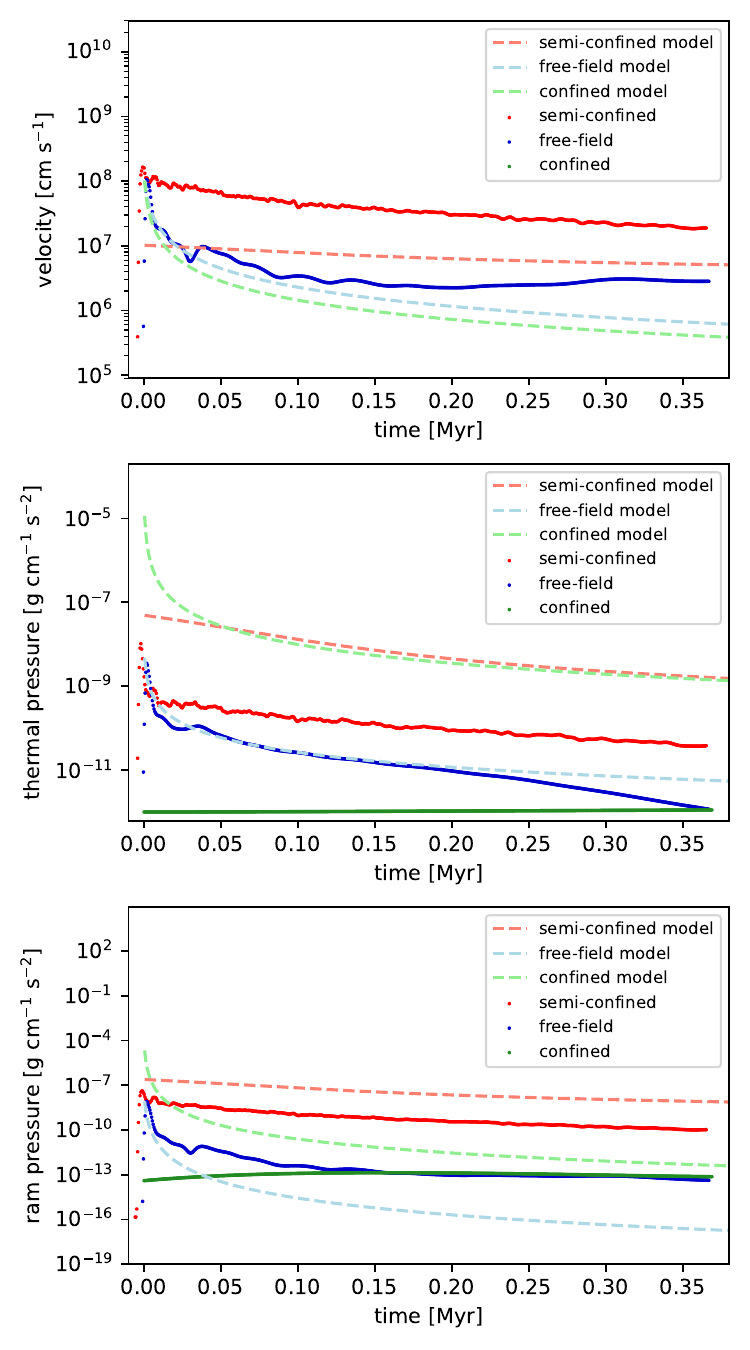}
    \caption{Same as Fig.~\ref{fig:outflow_adiabatic} but for the cooling run. The SN shock front did not cross the detector in the confined case.} 
    \label{fig:outflow_cooling}
\end{figure}

The plots for the runs with radiative cooling are shown in Fig.~\ref{fig:outflow_cooling}. It is apparent that, here, the free-field curves deviate significantly from their predictions obtained from the 1-D blast solutions. This is an inevitable consequence of the cooling, since it removes a significant portion of the energy from the SN after the shock thermalizes and subsequently terminates its Sedov phase. As seen from the second panel, the free-field thermal pressure begins to fall at around $0.20\ \mathrm{Myr}$ when radiative cooling starts to dominate. However, this still does not explain why the gas velocities level out (see first panel) instead of decaying like Fig.~\ref{fig:outflow_adiabatic} after the shock front has left. We suggest two other factors that are at play: the velocity profile of the ejecta particles that results in a more `extended' shock, or, the gas equilibrium temperatures from our heating and cooling implementations. A detailed explanation is provided in Appendix~\ref{appen:vel_cooling_ff}. 

Contrary to the above, the evolution of gas velocities in the semi-confined case agrees with the analytical model, with only a small systematic offset. One might question why the velocities are higher than that in the adiabatic runs when the shock has weakened. Recall that the velocities here are not necessarily an indicator of shock strength. The higher local velocity could be attributed to (a) the shock front is no longer highly supersonic and the cooled post-shock region is less well-modelled by the Rankine--Hugoniot conditions, which is meant to lower the post-shock velocities just like the free-field case, or (b) the channel size remained small (cf. Fig.~\ref{fig:sncavity_evol} and see Section~\ref{sec:outflow_chnlsize}).  

Setting aside the systematic offsets caused by overestimated densities, the thermal and ram pressures also show good agreement with their predicted evolutions. Thermal energy is lower than that in the adiabatic case as expected from cooling, nonetheless the analytical model still reproduces their decaying rate despite its adiabatic assumptions. It may only be the case if the cavity and channel(s) are sufficiently low in density, allowing the radiative cooling to be moderately suppressed. The results here again show that semi-confined SNe can induce a stronger perturbation to its local environment as long as the remnant within the cloud remains hot. 

Note also that in the confined case, no outflows were driven and the cavity did not expand beyond the detector location. Hence, the green curves in Fig.~\ref{fig:outflow_cooling} are \textit{not} measuring the feedback from the SN, but only the properties of the compressed cold gas in the cloud. 

\subsubsection{Multiple channels} \label{sec:outflow_nchnl}

\begin{figure}
    \includegraphics[width=3.2in]{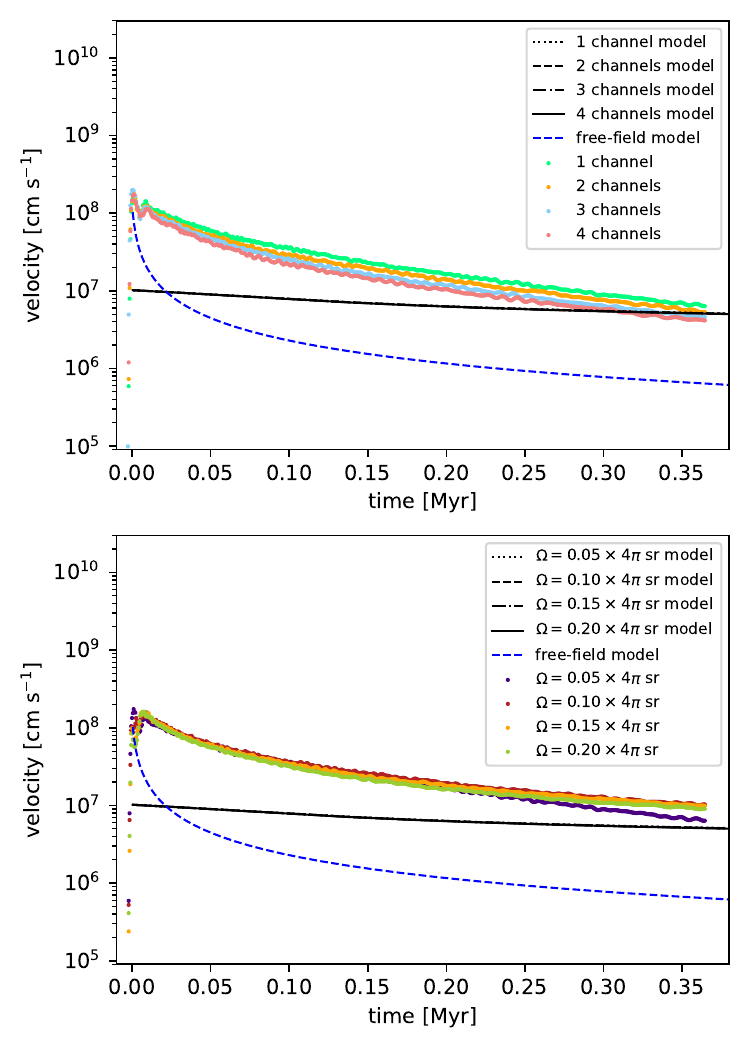}
    \caption{Velocity of outflow gas measured at the detector location from the adiabatic runs compared to the predictions from the analytical model (black). \textit{Top}: Semi-confined simulations with one to four channels carved around the progenitor, each with solid angle $\Omega = 0.05\times4\pi\ \mathrm{sr}$. The green curve is identical to the red semi-confined curve in Fig.~\ref{fig:outflow_adiabatic}. \textit{Bottom}: Semi-confined simulations with channel sizes $\Omega = 0.05\times4\pi\ \mathrm{sr}$, $\Omega = 0.10\times4\pi\ \mathrm{sr}$, $\Omega = 0.15\times4\pi\ \mathrm{sr}$ and $\Omega = 0.20\times4\pi\ \mathrm{sr}$. The purple curve is identical to the red semi-confined curve in Fig.~\ref{fig:outflow_adiabatic}.} The free-field curve is shown for reference. 
    \label{fig:mulvent_adiabatic}
\end{figure}

\begin{figure}
    \includegraphics[width=3.2in]{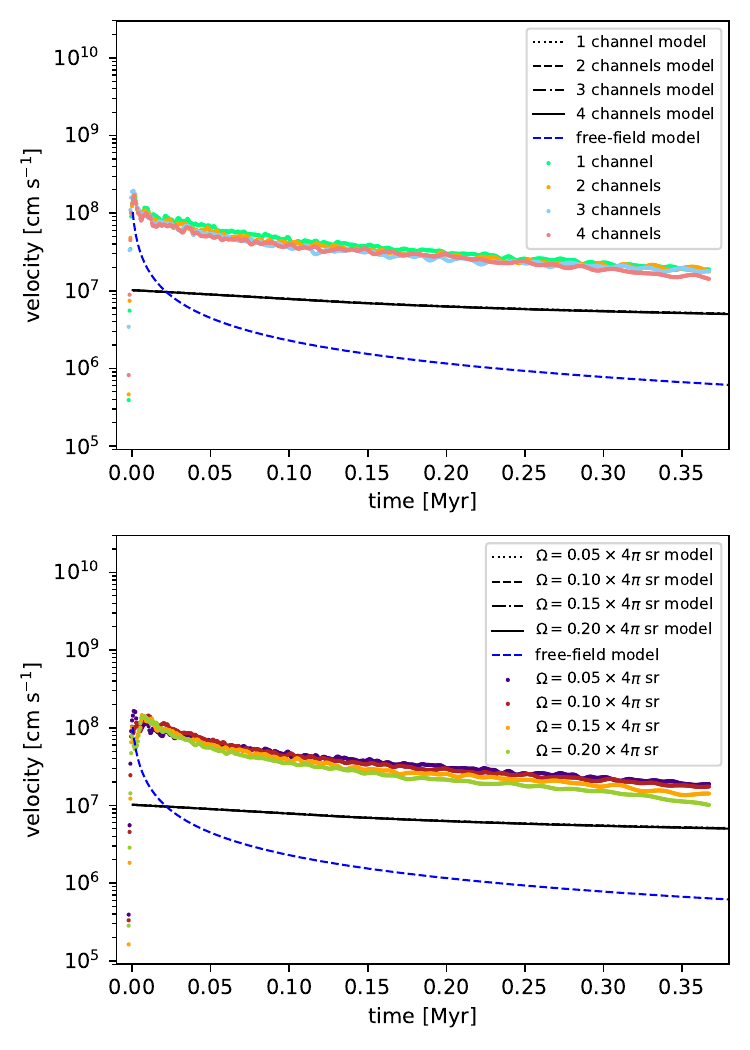}
    \caption{Same as Fig.~\ref{fig:mulvent_adiabatic} but for the cooling runs. \textit{Top}: Varying number of channels. The green curve is identical to the red semi-confined curve in Fig.~\ref{fig:outflow_cooling}. 
    \textit{Bottom}: Varying channel sizes. The purple curve is identical to the red semi-confined curve in Fig.~\ref{fig:outflow_cooling}.} 
    \label{fig:mulvent_cooling}
\end{figure}

It is also possible to assess the effect of increasing the number of channels. In the top panels of Fig.~\ref{fig:mulvent_adiabatic} and Fig.~\ref{fig:mulvent_cooling} we plot the evolution of gas velocities at the detector in four sets of simulations, in which the molecular clouds have one to four identical channels respectively, adiabatic and with cooling. Each channel is carved with the same method described earlier, along four different Cartesian axes relative to the progenitor. The detector is placed in front of one of the channels. Because the setup is geometrically symmetric, the same results would be obtained independent of which channel the detector is placed. 

We also plot the predictions calculated with the analytical models. Their results suggest that a small increase in size of the vent imposes negligible effect on the pressure relief. This prediction seems counter-intuitive, that the energy and pressure reduction within the cavity is independent of the number of channel(s) present around the bubble. Indeed, there is a possibility that this is merely a consequence of our small-vent assumption used in the derivations. However, the results obtained from simulations appear to suggest that the model predictions are in fact correct, despite that the channel sizes are expanding and the outflows are interacting with each other. All curves remained close to the one-channel curve in both the adiabatic run and the cooling run; only small differences are seen due to the higher mass loss rate. The results show that our simple analytical model is capable of largely reproducing the behaviour of outflows from bubble leakages.

\subsubsection{Bigger channels} \label{sec:outflow_chnlsize}

As mentioned in Section~\ref{sec:semiconf_model_outflow}, our analytical model does not distinguish between a large channel and multiple smaller channels that cover the same surface area on the cloud. To test whether or not the model breaks down when the outflow departs from being a streamline, we also performed simulations of semi-confined SNe with opening angles of $\Omega = 0.10\times4\pi\ \mathrm{sr}$, $\Omega = 0.15\times4\pi\ \mathrm{sr}$ and $\Omega = 0.20\times4\pi\ \mathrm{sr}$. These solid angles correspond to those in Section~\ref{sec:outflow_nchnl} but combined into a single channel. The results are plotted in the bottom panels of Fig.~\ref{fig:mulvent_adiabatic} and Fig.~\ref{fig:mulvent_cooling}. 

Surprisingly, the curves remain very similar to run with channel size $\Omega = 0.05\times4\pi\ \mathrm{sr}$, and overall exhibit a very similar behaviour to their multi-channel counterparts. More importantly, the plots reveal that the adiabatic semi-confined run presented in Fig.~\ref{fig:outflow_adiabatic} (equivalent to the purple curve in Fig.~\ref{fig:mulvent_adiabatic}) is, in fact, an anomaly. Its velocity falls at a higher rate compared to the rest and is against the trend seen in Fig.~\ref{fig:mulvent_cooling}. We believe this is due to the size of the channel, which had expanded significantly upon collision with the strong adiabatic shock. To the contrary, those with larger initial channel sizes did not experience this impact, because their low-density pathways are already sufficiently broad that the energy could `leave quietly' without causing much disruption to the channel walls. This result is in line with the findings of \citet{lucas20}, who reported on a similar behaviour in turbulent molecular cloud simulations. With a relatively static channel size and a stable rate of energy release, the local velocities remain in agreement with their predicted values from the analytical model (cf. the red, orange, green curves in bottom panel of Fig.~\ref{fig:mulvent_adiabatic}). What it implies is that the steep drop in velocity seen in Fig.~\ref{fig:outflow_adiabatic} likely originates from the rapid energy loss from the cavity due to channel expansion. 

Overall, the results described in Section~\ref{sec:outflow_nchnl} and Section~\ref{sec:outflow_chnlsize} have crucial implications on how the impact of a SN varies with the density structure of its environment. It could mean that the outflows through different small gaps on the feedback-driven shells share similar kinematic properties, regardless of the fractal dimension of its host cloud.

\subsection{Energy and Momentum deposition} \label{sec:energy_momentum}

\begin{figure*}
    \includegraphics[width=6.7in]{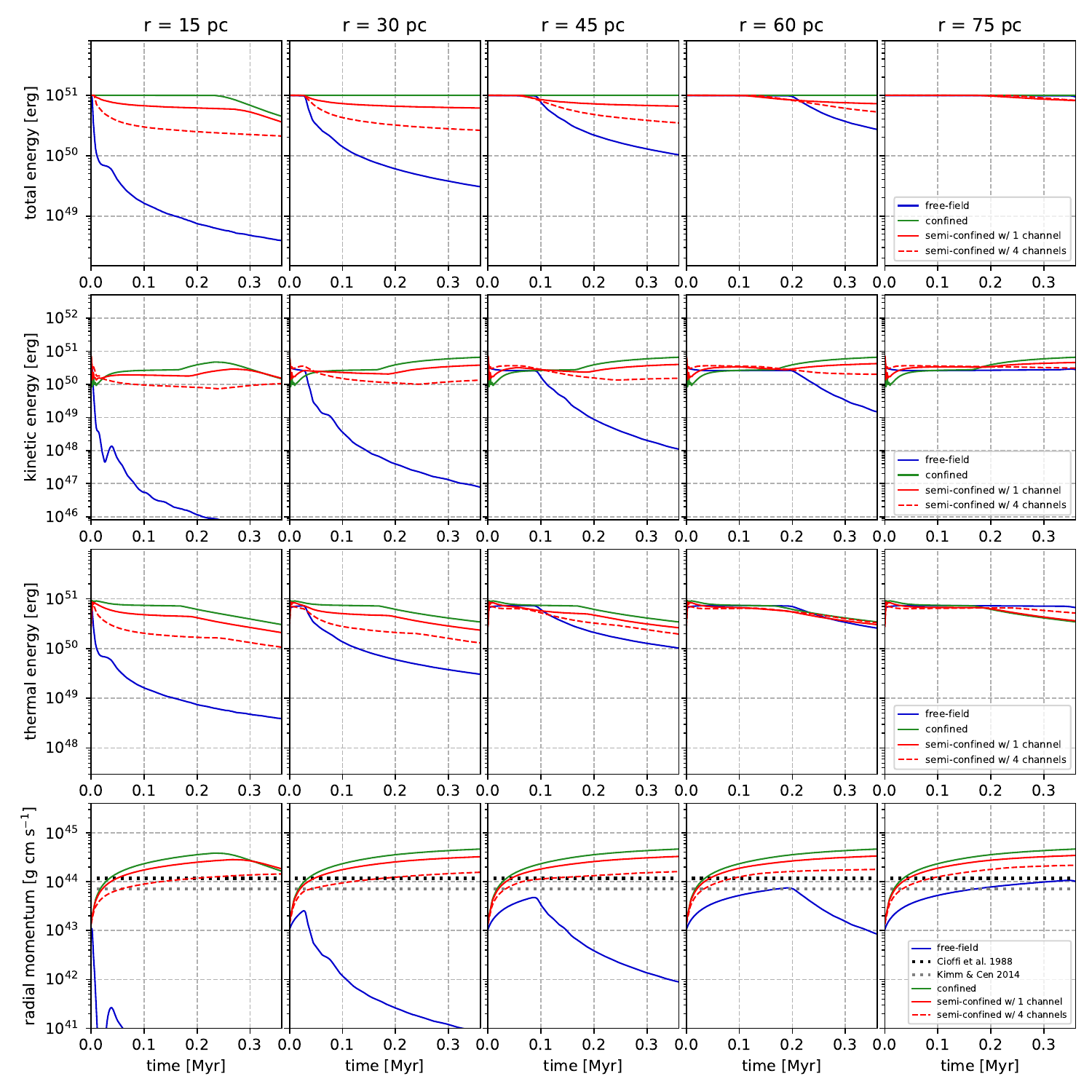}
    \caption{\textit{Left to right}: Energy and momentum contained within a radius of $15\ \mathrm{pc}$, $30\ \mathrm{pc}$, $45\ \mathrm{pc}$, $60\ \mathrm{pc}$ and $75\ \mathrm{pc}$ around the progenitor in the adiabatic run. \textit{First row}: The total amount of energy in the free-field run (blue), confined run (green), the semi-confined run with one channel (red solid), and the semi-confined run with four channels (red dashed). \textit{Second row}: The amount of energy in kinetic form. \textit{Third row}: The amount of thermal energy, including the thermalized kinetic energy in the SN shocks. \textit{Fourth row}: The total radial momentum carried by the particles. The analytical expressions for terminal momentum formulated by \citet{cioffi88} and \citet{kimmcen14} are indicated by the black and grey horizontal dotted lines respectively. }  
    \label{fig:energy_momentum_adiabatic}
\end{figure*}

This section examines the coupling efficiency in energy and momentum between the SNe and their surrounding medium. Fig.~\ref{fig:energy_momentum_adiabatic} plots the total energy, kinetic energy, thermal energy and radial momentum contained within a radius of $15\ \mathrm{pc}$, $30\ \mathrm{pc}$, $45\ \mathrm{pc}$, $60\ \mathrm{pc}$ and $75\ \mathrm{pc}$ around the progenitor in the adiabatic run\footnote{The initial amount of thermal energy contained within the measuring radii are as follows. Confined: $3.13\times10^{49}\ \mathrm{erg}$ (15 pc) to  $3.32\times10^{49}\ \mathrm{erg}$ (75 pc). Semi-confined: $2.98\times10^{49}\ \mathrm{erg}$ (15 pc) to  $3.16\times10^{49}\ \mathrm{erg}$ (75 pc). Free-field: $1.52\times10^{46}\ \mathrm{erg}$ (15 pc) to  $1.81\times10^{48}\ \mathrm{erg}$ (75 pc). We consider them insignificant compared to the injected SN energy. }. We include results for the free-field case, the confined case, and two semi-confined cases with one and four channels respectively.

It is apparent from Fig.~\ref{fig:energy_momentum_adiabatic} that despite the total amount of energy released on large-scales ($> 75\ \mathrm{pc}$) are similar across all cases, the semi-confined SNe are much more capable of retaining a large fraction of its injected kinetic energy within a radii of $\sim 60\ \mathrm{pc}$. Even with four channels, the amount of energy retained does not differ significantly. This is in contrast with the free-field case, in which the shock energy has rapidly reached $\sim 75\ \mathrm{pc}$ within $0.2\ \mathrm{Myr}$, leaving little energy deposited at smaller distances. Here, the confined case appears to be the best at retaining energy within $\sim 30\ \mathrm{pc}$. The total energy remained constant, and local drops in thermal energy were in turn converted into kinetic energy, possibly via the PdV work of the gas. However, we shall see that this is no longer the case when radiative cooling is being incorporated. 

We also measure the total radial momentum carried by the particles, and compare the results to analytical expressions in literature that describe the \textit{terminal momentum} of spherically symmetric SNe. Towards the end of their pressure-driven snowplough (PDS) phase where radiative cooling has removed the majority of its interior thermal energy, the shell momentum begins to plateau. \citet{cioffi88} showed that the maximum achievable momentum of a SN blast can be described by
\begin{equation}
    \bar{\mu}_\mathrm{final} = 4.8 \times 10^5 \frac{E_{51}^{13/14}}{\zeta_m^{3/14} n_H^{1/7}} \qquad \mathrm{[M_\odot \ km \ s^{-1}]}
    \label{eq:cioffi_terminal_momentum}
\end{equation}
\citep[cf. ][equation 4.7]{cioffi88}, where $E_{51}$ denotes SN energy in units of $10^{51}\ \mathrm{erg}$, and $\zeta_m \equiv Z/Z_\odot$ is the metallicity factor. Here, we set $\zeta_m = 1$, consistent with the assumption of the \citet{joungmaclow06} cooling curve. To allow for comparisons with our free-field model, we take $n_H = \rho_\mathrm{amb}/m_H$. 

\begin{figure*}
    \includegraphics[width=6.7in]{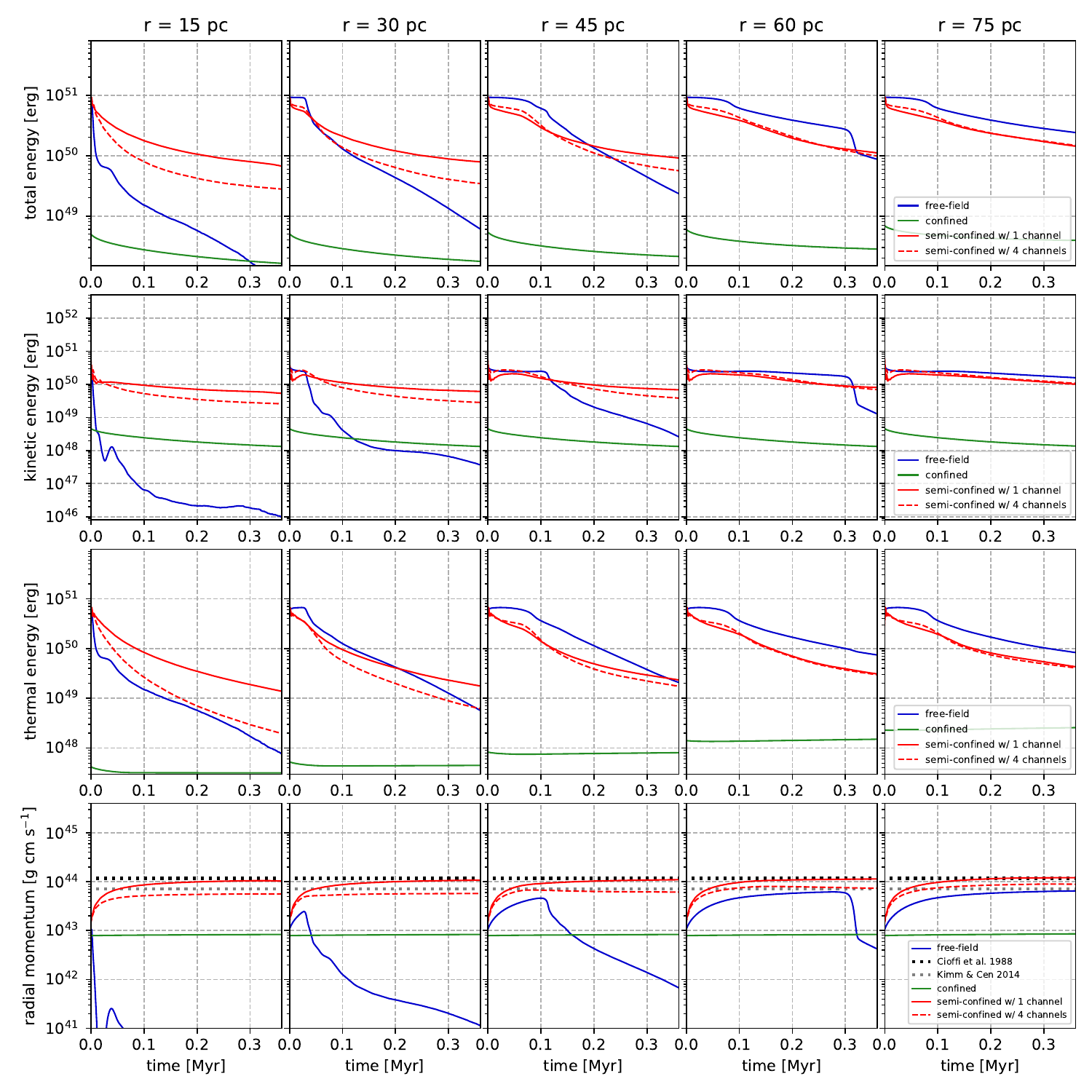}
    \caption{Same as Fig.~\ref{fig:energy_momentum_cooling} but for the cooling runs. }
    \label{fig:energy_momentum_cooling}
\end{figure*}

Similarly, \citet{kimmcen14} also formulated an expression for the terminal momentum based on the results of \citet{thornton98}, who showed that the momentum input from SNe decreases with metallicity of the ambient medium. Their expression reads:
\begin{equation}
    \bar{\mu}_\mathrm{final} = 3.0 \times 10^5 \frac{E_{51}^{16/17}}{n_H^{2/17}} \mathrm{max}[\zeta_m,0.01]^{-0.14} \quad \mathrm{[M_\odot \ km \ s^{-1}]}
    \label{eq:kimmcen_terminal_momentum}
\end{equation}
\citep[cf.][equation A4]{kimmcen14}. The term involving $\zeta_m$ is constrained because the metallicity-dependence vanishes below $0.01\ Z_\odot$. Analytical expressions like equations~(\ref{eq:cioffi_terminal_momentum}) and (\ref{eq:kimmcen_terminal_momentum}) are of crucial importance for the mechanical/kinetic feedback type of sub-grid models \citep[e.g.][]{hopkins11,hopkins13,rosdahl17}, particularly if the simulation is unable to resolve the Sedov--Taylor phase and require a direct injection of the momentum that is left from the radiative snowplough phases. We plot the two equations in the bottom panels of Fig.~\ref{fig:energy_momentum_adiabatic} and Fig.~\ref{fig:energy_momentum_cooling} for comparison. 

Similar to the kinetic energy plots, in Fig.~\ref{fig:energy_momentum_adiabatic} we see that the confined and semi-confined cases deposit the majority of their momentum within a distance of $\sim 15\ \mathrm{pc}$ from the progenitor. This distance is significantly shorter than that of the free-field, whose terminal momentum is reached only beyond $\sim 75\ \mathrm{pc}$. Furthermore, whilst the free-field case correctly approached the terminal momentum of \citet{cioffi88} after 0.3 Myr, the momentum deposited by the confined and semi-confined cases exceeded it within 0.05 Myr. This is likely owing to the collisions with the cavity walls, during which momentum deposition is immediately maximized. Even in the run with four channels, the amount of momentum deposited is slightly lowered due to reduced collisions, nonetheless it remains higher than that of the free-field case as well as the analytical predictions. 

We now consider the effect of cooling. Radiative cooling is the dominant constraint on the energy and momentum coupling efficiency between the SNe and its surrounding medium. Fig.~\ref{fig:energy_momentum_cooling} shows the results. In the fully-confined case, the injected energy was immediately thermalized and was almost completely lost to cooling upon collision. It was unable to push against the cavity walls, as previously seen in Fig.~\ref{fig:outflow_cooling}. Hence, the low kinetic energy and momentum shown here only reflect the minor disturbances in the cloud caused by the collision. Since no outflows were driven, we do not further discuss on confined models. 

Comparing the thermal energies in the free-field and the semi-confined cases, they decay at a very similar rate. Since the cooling in our algorithm is only determined by gas density, this consistency is achieved only if the cavity density is sufficiently low (cf. Section~\ref{sec:setup}). Otherwise, the cavity would rapidly cool off and cease to drive outflows out of the channel. As such, for there to be any meaningful comparison, we must ensure that the free-field and the semi-confined SN are subjected to similar amounts of radiative cooling. 

Fig.~\ref{fig:energy_momentum_cooling} also shows that semi-confined SNe are much better at retaining kinetic energy at short distances from the progenitor, inducing a higher dynamical perturbation to their local environment. However it does not necessarily mean that the size-scale of the SN remnant is reduced. We see from Fig.~\ref{fig:energy_momentum_cooling} that the amount of kinetic energy deposited becomes on par with the free-field case only until $\sim 75\ \mathrm{pc}$, meaning the two have similar impact radii. We shall also see in Section~\ref{sec:turbulence} that the shock front in the semi-confined case can reach equally large distances as in the free-field case. 

More importantly, the momentum deposition from the semi-confined SN peaked within $15\ \mathrm{pc}$, shortly after the collision with the cavity walls. Its terminal momentum precisely matches the prediction of \citet{cioffi88}. Whether or not this is a coincidence warrants further investigation. With four channels, the amount of momentum deposited remains similar, and shows good agreement with the prediction of \citet{kimmcen14}. On larger scales, the difference between the semi-confined and the free-field is less prominent compared to the adiabatic runs, but the semi-confined case deposits slightly more radial momentum overall. Crucially, it shows that the amount of momentum imparted locally around the progenitor can be severely underestimated if partial confinement effects are not taken into account. It also shows that the time-scale of momentum coupling could be longer than previously expected, since deposition could commence immediately after the SN explosion, and continue until the remnant merges with the ISM.

Additionally, these results supplement to the findings of \citet{rosen14}, who suggested that the large fraction of missing energy within feedback-driven bubbles is attributed to the presence of holes on the porous shell. The authors showed that the shell cover fraction $C_f$ should be around the range 0.36--0.95 in order to explain their observed luminosities. Similarly, \citet{dale14} found that the leakage factor lie mostly at around 0.6--0.8, though both studies only accounted for pre-SN feedback. Our findings here suggest that $C_f$ may need to be much lower than 0.8, which otherwise the confinement effect would still dominate and prevent the bubble from losing the required amount of energy.

\subsection{Turbulence driving} \label{sec:turbulence}

We showed in the previous section that semi-confined SNe are capable of sustaining a higher level of kinetic input at short distances to its progenitor compared to spherical expansions. Since SNe are one of the primary drivers of turbulence in the ISM \citep[e.g.][]{elmegreenscalo04,balsara04,maclow04,bacchini20}, we further investigate the semi-confined SNe’s mechanical input by assessing their ability to induce compressive and solenoidal turbulent motions. 

Shocks expanding into a low-density circumstellar medium naturally produces helicity in the ISM. When the shell of SN ejecta decelerates after interacting with the shocked ambient gas, the ejecta materials at the interface can become unstable to Rayleigh--Taylor instability and subsequently develop the so-called Rayleigh--Taylor fingers that protrude into the surrounding medium \citep[][]{jun96,blondinellison01}. As the fingers grow, the shear at the contact boundaries gives rise to secondary Kelvin--Helmholtz instability, generating vorticities along the fingers \citep[e.g.][]{bucciantini04}. Such eddies might in turn enhance the growth of the fingers by supplying extra rotational kinetic energy \citep[][]{jun96}. Even in the absence of hydrodynamic instabilities, strong shocks that interact with a clumpy turbulent medium also naturally generate velocity dispersions \citep[][]{dobbsbonnell08,bonnell06,bonnell13} and helicity fluctuations in the post-shock regions \citep[][]{balsara01,balsara04}. A similar finding has been reported in \citet{padoan16}, that the solenoidal component in the turbulence must be greater than the compressive if SN interacts with a medium whose density gradients are in random directions. 

Though turbulence is not being added in our simulations as a starting condition, our semi-confined SN model ubiquitously involves shock collisions with cavity walls. We thus expect a larger amount of shear and vortices to develop in their outflows compared to free-fields (see the plumes shown in Fig.~\ref{fig:sncavity_evol}). As the fluid moves through the channels, it may be also subjected to Kelvin--Helmholtz instability that generates small-scale eddies. A higher helicity in the flow implies a stronger solenoidal component in its velocity field. Solenoidal turbulences are known for their ability to support against gravitational collapse and regulate star formation \citep[e.g.][]{rani22}. 

To compare the amount of vorticity in the semi-confined and the free-field cases, we decompose their velocities to plot the kinetic energy spectra. Our method works by, first, constructing a cubical Cartesian grid that covers a spatial volume in the simulation. The grid vertices serve as interpolation points where we obtain the fluid's velocity from the neighbouring particles. The length of the grid and the spacing between the vertices hence define our resolution limits. The velocity field is subsequently passed into a Helmholtz--Hodge decomposition algorithm to separate the total velocity into solenoidal (divergence-free) and compressive (curl-free) components. The velocity cubes for each component are then passed into a fast Fourier transform algorithm to produce their kinetic energy spectra in wave number space. We also emphasize that, since we are interested only in the generation of local kinetic motions in the post-shock regions, the grid does not necessarily cover the shock front. 

\begin{figure}
    \includegraphics[width=3.3in]{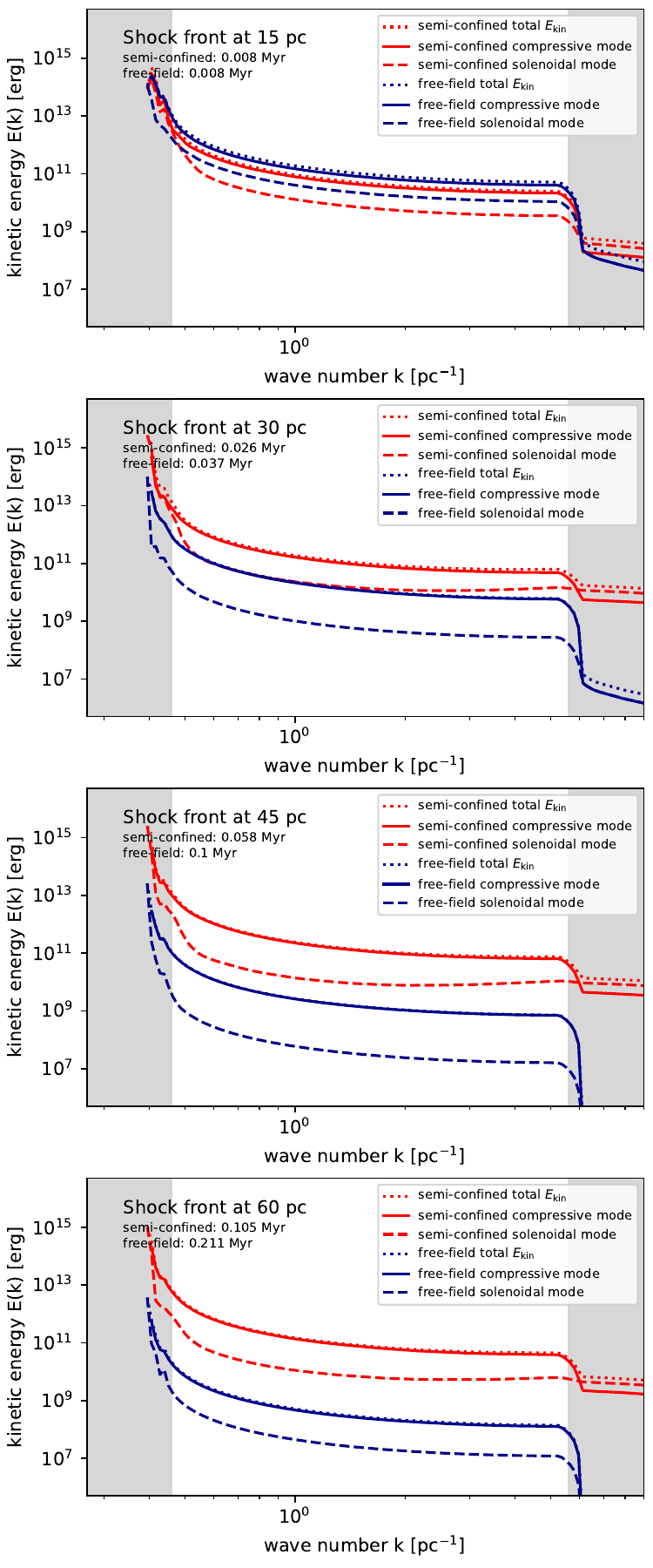}
    \caption{\textit{Top to bottom}: Kinetic energy spectra of the adiabatic runs comparing semi-confined (red) to free-field (blue) when their shock fronts both reach a distance of approximately $15\ \mathrm{pc}$, $30\ \mathrm{pc}$, $45\ \mathrm{pc}$ and $60\ \mathrm{pc}$ from the progenitor. Each plot shows the total kinetic energy (dotted), its compressive component (solid), and its solenoidal component (dashed). Note the time difference between the free-field and the semi-confined curve being compared in each panel. The greyed-out regions indicate the resolution limits and may be disregarded. }
    \label{fig:turbspec_adiabatic}
\end{figure}

Fig.~\ref{fig:turbspec_adiabatic} shows the spectra extracted from the adiabatic runs. The grid has $200^3$ interpolation points with a length of $16\ \mathrm{pc}$ (radius of $8\ \mathrm{pc}$), centred upon the x-axis at $20\ \mathrm{pc}$ away from the progenitor, such that it lies \textit{just} beyond the cloud. We compare semi-confined to free-field at time instances where their shock fronts are at similar distances from the progenitor, hence a similar driving scale. The resolution limits constrained by the grid spacing and its full length are greyed out to indicate the parts of the curve to be neglected. We compare their spectra at shock front radii of $15\ \mathrm{pc}$, $30\ \mathrm{pc}$, $45\ \mathrm{pc}$ and $60\ \mathrm{pc}$ respectively. The semi-confined curves are plotted in red and those of free-field are in blue, each with their total kinetic energy and their solenoidal and compressive components displayed. The peaks seen at large length scales indicate a strong driving. Note however that the location of this peak is likely limited by the length of our grid, hence we consider here only their spectra intensities rather than their precise wave numbers.

As seen from the first panel in Fig.~\ref{fig:turbspec_adiabatic}, the kinetic energy spectra of the free-field and the semi-confined are initially alike, though the solenoidal component of the semi-confined appears slightly weaker. Once the shock front reaches $30\ \mathrm{pc}$, the kinetic energy of the free-field begins to fall, in agreement with the findings from Fig.~\ref{fig:energy_momentum_adiabatic}. Meanwhile, in the semi-confined case, its solenoidal component rises slightly at large $k$. This contradicts the typical spectrum of a turbulent cascade and likely indicates the presence of small-scale eddies, which could have arose \textit{directly} from the collisions and shear flows in the channel. As the shock evolves to $60\ \mathrm{pc}$, the local kinematic perturbation continues to weaken in the free-field case, particularly in its solenoidal components. This is in contrast with the semi-confined case, in which the spectrum seems to have maintained its high intensities throughout the evolution. It indicates that semi-confined progenitors can induce more sustained turbulent flows in its local environment. 

\begin{figure}
    \includegraphics[width=3.3in]{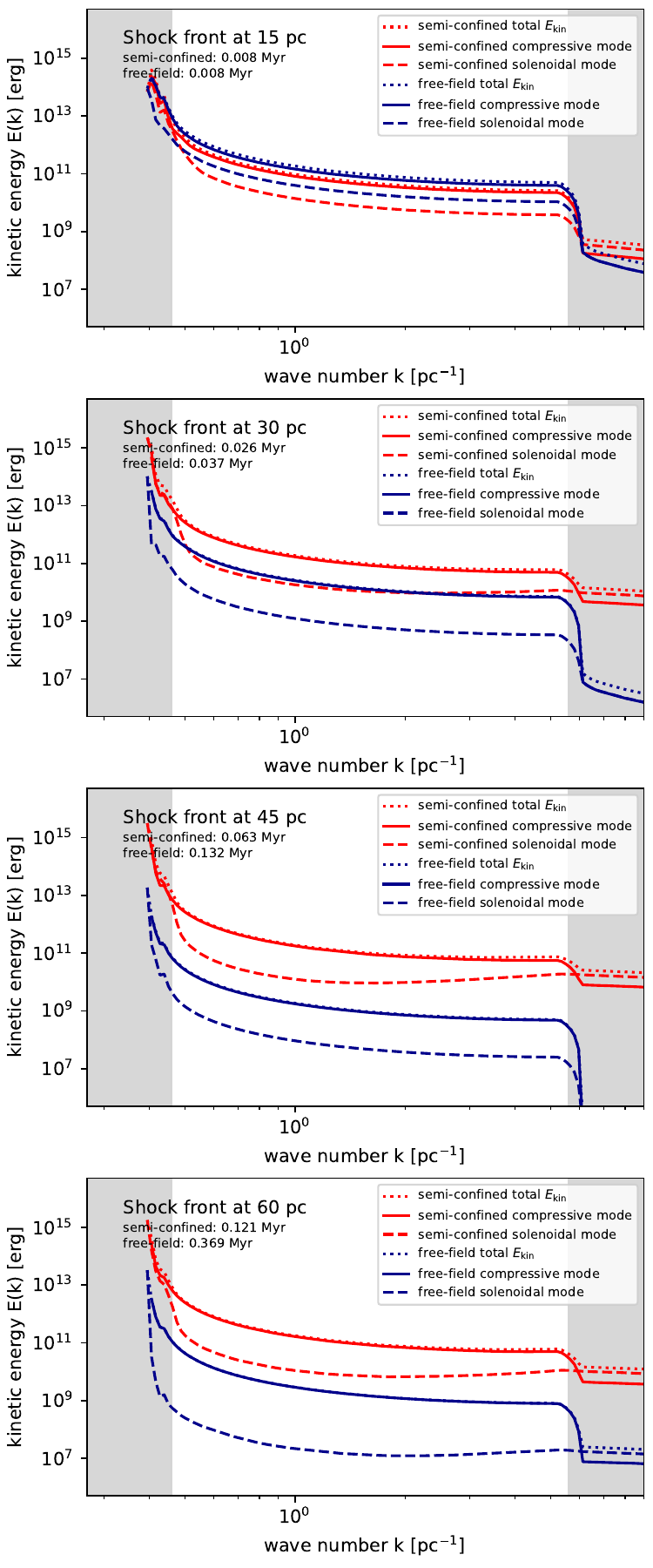}
    \caption{Same as Fig.~\ref{fig:turbspec_adiabatic} but for the cooling runs. }
    \label{fig:turbspec_cooling}
\end{figure}

We repeat the same measurements with the cooling runs. Their kinetic energy spectra are presented in Fig.~\ref{fig:turbspec_cooling}. It can be seen that they exhibit a very similar behaviour to the adiabatic runs. The rise in solenoidal component at larger values of $k$ in the semi-confined case is even more prominent compared to Fig.~\ref{fig:turbspec_adiabatic}. Indeed, it is no surprise that cooling would give rise to fluid instabilities and introduce higher level of vorticity in the outflows as it reduces the gas thermal pressure. This may be analogous to the idea discussed in \citet{kuiper12}, that radiation-driven cavities around young massive stars are prone to Rayleigh--Taylor instabilities when the radiative pressure support is underestimated.

At later times, when the shock front has reached $60\ \mathrm{pc}$, small-scale eddies are seen to emerge in the free-field case as well. The shell expansion is no longer driven by strong thermal pressure, and the gas particles begin to relax to their equilibriums. It is likely that this process could have induced chaotic motions within the remnant that in turn gave rise to the development of fluid instabilities.  

Nonetheless, the results show that the spacings of our measuring grid is sufficiently small for capturing the eddies in the SN-driven flows through porous shells. We conclude from Fig.~\ref{fig:turbspec_adiabatic} and Fig.~\ref{fig:turbspec_cooling} that semi-confined SNe can induce and maintain a higher level of solenoidal perturbation to its surrounding ISM, owing to the vorticities generated by the plumes in its outflows. A smaller grid may be able to further reveal the details in the kinematic properties of this plume as it interacts and mixes with the warm envelope gas.

\section{Discussions and Summary} \label{sec:conclusion}

In this paper, we presented a theoretical model of a SN explosion that is partially confined within a cavity carved by its earlier stellar feedback. Since past studies revealed that SN energy often leave their natal cloud preferentially through low-density channels, we investigated how the SN output in this scenario would differ to that from a typical 1-D blast model. We showed that our analytical model successfully predicts the outflow velocities measured from 3-D numerical simulations, even for the runs with radiative cooling and those with multiple or larger channels around the porous cavity shell. We also studied the gas properties of the outflow, its energy and momentum deposition, and the driven turbulence. Our results showed that semi-confined SNe impose stronger dynamical impact on the smaller scales, as they deposit a larger amount of kinetic energy and momentum at radii closer to the progenitor. The coupling timescales are extended thanks to the slow pressure relief from the cavity. From the kinetic energy spectra, the solenoidal turbulence driven by semi-confined SN are also seen to be amplified at smaller length-scales via the development of plumes. 

A few key interpretations can be drawn from our results: 
\begin{enumerate}
    \item Large-scale simulations that rely on Sedov--Taylor-like solutions as sub-grid models of SN might underestimate the amount of local kinetic energy and momentum deposition.  
    \item Peculiar velocity profiles observed around highly asymmetrical SN remnants can be explained by outflows through the porous shells, whose velocities can be much higher than if the SN shock is expanding isotopically. 
    \item Outflows through gaps on the bubble may be treated as individual streamlines. It provides good estimates for the outflow velocities and pressure just using simple fluid equations that govern adiabatic incompressible flows. 
    \item Terminal momentum of SNe can be reached at much smaller length-scales and shorter time-scales if it is partially confined.
    \item 1-D SN models that account for the collision with feedback-driven cavity walls but neglect the outflows through the gaps could overestimate the amount of cooling experienced by the remnant.
    \item In order to explain the energy or pressure loss from feedback-driven bubbles \citep[e.g.][]{rosen14,lopez14}, the cover fraction of the shell must be sufficiently small. We showed in our simulations that even a total opening area of 20\% of the spherical shell surface does not lead to significant pressure loss from the cavity (cf. Fig.~\ref{fig:mulvent_cooling}). 
    \item That semi-confined SNe drive stronger solenoidal turbulence at regions near the molecular cloud may indicate a higher ability to regulate star formation in their local environments. 
    \item An extended kinetic energy coupling timescale with its natal molecular cloud can imply that a lower SN rate may be sufficient to reproduce the observed local star formation efficiencies. 
\end{enumerate}

The differences between a semi-confined model and a spherically symmetric model have plenty of implications on the large-scale evolutions. We argue that it is necessary to account for the partial confinement effects and the collimated outflows from SNe in the 1-D sub-grid models. This may be done via parametrization accompanied with statistical techniques. Additionally, since most of the current SN models can already account for non-homogeneous ambient ISM, incorporating density fluctuations into our models' background medium before carrying out the comparisons may provide more useful information on the limitations of these 1-D models. Of course, a major shortcoming of our study is the overly solid cavity wall, which was created only to replicate the analytical model. Turbulent mixing at the interface between SN remnant and the shell could have a significant impact on the energy dissipation \citep[e.g.][]{mackey15}. As such, studying embedded SNe in more realistic turbulent molecular cloud environments should be the next step to verify our findings. 

On a final note, it may be beneficial to extend this study by incorporating magnetic fields. Spherical shells driven by expanding H {\scshape ii} regions that collide with ISM materials are capable of compressing magnetic fields and bending the field lines along the dense shells \citep[e.g.][]{pattle23,tahani23}. These tangential fields may suppress shell fragmentation and resist the formation of radiation-driven pillars or protrusions from the H {\scshape ii} regions \citep[][]{hennebelleinutsuka19}. It may be of interest to study whether or not magnetic fields can hinder SN outflows from leaky feedback bubbles.

\section*{Acknowledgements}

We thank our reviewer Ben Keller for the very helpful comments and suggestions that significantly improved the paper. CSCL is also grateful for the input from Ben and Craig Yanitski on the project idea. CSCL thank Daniel Seifried and colleagues from Theoretical Astrophysics group Cologne for their comments regarding this work. This project was supported by the STFC training grant ST/W507817/1 (Project reference 2599314). The simulations were performed using the HPC Hypatia, operated by University of St Andrews. The SPH figures in this paper were created using the python package {\scshape sarracen} \citep[][]{sarracen23}. The Helmholtz--Hodege decomposition algorithm was developed based on the python code {\scshape helmholtz} \citep[][]{shi18}. The algorithm for generating kinetic energy spectra was developed based on the python code {\scshape tkespec} \citep[][]{saad14}. 

\section*{Data Availability}

The model calculation results and simulation outputs underpinning this publication can be accessed at: \url{https://doi.org/10.17630/20d675b9-9ce7-4a1e-b4fe-2cbc34a99b0e}. Newest versions of the simulations codes are publicly available at \url{https://github.com/Cheryl-Lau/phantom}.



\bibliographystyle{mnras}
\bibliography{ref_list}




\appendix

\section{Equilibrium temperatures} \label{appen:equil_temp}

\begin{figure}
    \includegraphics[width=3.35in]{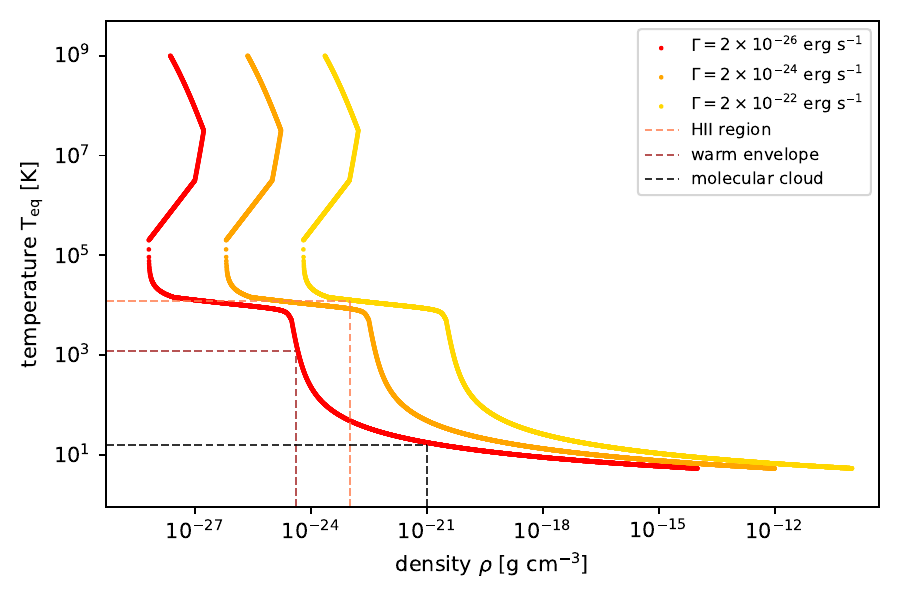}
    \caption{Equilibrium temperatures $T_\mathrm{eq}$ as a function of gas local density $\rho$ obtained by solving the thermal equilibrium equation, $n_H\Gamma - n_H^2\Lambda(T) = 0$, where $\Lambda(T)$ is defined by a cooling curve, and heating term is $\Gamma = 2 \times 10^{-26}\ \mathrm{erg\ s^{-1}}$ (red), $\Gamma = 2 \times 10^{-24}\ \mathrm{erg\ s^{-1}}$ (orange) and $\Gamma = 2 \times 10^{-22}\ \mathrm{erg\ s^{-1}}$ (yellow). The red curve shows equilibrium temperatures when the gas is only subjected to background FUV heating, and yellow curve shows the equilibrium temperatures under stellar photoionization heating. Dashed lines mark the temperatures of the cloud (black), the envelope (brown) and the H {\scshape ii} region (light brown), whose densities are given in Section~\ref{sec:setup}. }
    \label{fig:teq_solutions}
\end{figure}

To efficiently update the internal energies with the implicit method (equation~\ref{eq:implicit_new_u}), we pre-compute and tabulate the equilibrium temperatures for a range of particle densities $\rho$, then interpolate from this table during runtime. Fig.~\ref{fig:teq_solutions} plots the solution temperatures $T_\mathrm{eq}$ that satisfy the thermal equilibrium equation $n_H\Gamma - n_H^2\Lambda(T) = 0$, where the background heating is $\Gamma = 2\times10^{-26}\ \mathrm{erg\ s^{-1}}$ \citep[][]{koyamainutsuka02} and the cooling $\Lambda(T)$ is governed by the cooling curve from \citet[Fig. 1]{joungmaclow06}. Note that $n_H = \rho/m_H$. The high-density regime in Fig.~\ref{fig:teq_solutions} is akin to the equilibrium temperature plot in e.g. \citet{bonnell13}, Fig. 2. The low-density regime, however, is multi-valued, meaning there can be more than one thermal equilibrium at certain gas densities. The method to identify the equilibrium to which the particle should approach is detailed in \citet{lau25}.

We also plot in Fig.~\ref{fig:teq_solutions} the solutions for $\Gamma = 2\times10^{-24}\ \mathrm{erg\ s^{-1}}$ and $\Gamma = 2\times10^{-22}\ \mathrm{erg\ s^{-1}}$ to illustrate the effect of increased heating. The latter is roughly equal to the stellar photoionization heating rate for cloud densities of around $10^{-21}\ \mathrm{g\ cm^{-3}}$ \citep[cf.][Fig. 9]{lau25}. Hence, the yellow curve governs the equilibrium temperatures within the H {\scshape ii} regions. For $\rho_\mathrm{HII} = 10^{-23}\ \mathrm{g\ cm^{-3}}$, its temperatures is at around $10^4\ \mathrm{K}$ (dashed light brown). Elsewhere in the simulation, the gas temperatures follow the red curve. Thus, the cloud with density $\rho_\mathrm{cloud} = 10^{-21}\ \mathrm{g\ cm^{-3}}$ is at $10\ \mathrm{K}$ (dashed black), and the warm envelope with $\rho_\mathrm{env} = 4\times10^{-25}\ \mathrm{g\ cm^{-3}}$ is at $10^3\ \mathrm{K}$ (dashed brown).

\section{Free-field outflow gas velocity at detector in cooling runs} \label{appen:vel_cooling_ff}

\begin{figure}
    \includegraphics[width=3.35in]{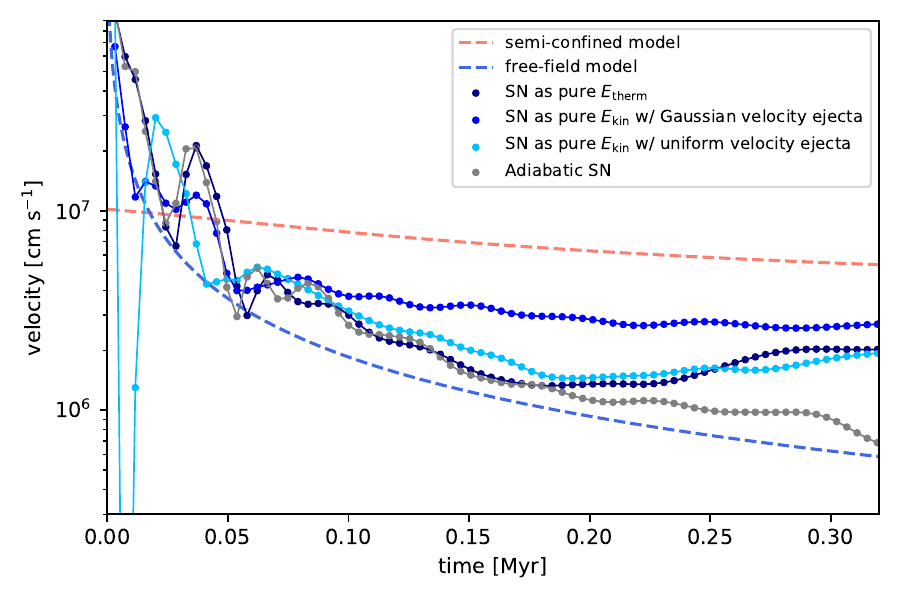}
    \caption{Time-evolution of gas velocity at the detector location in the cooling run for a free-field SN that is injected as pure thermal energy (navy), one that is injected as pure kinetic energy with uniform radial velocities assigned to the ejecta particles (light blue), and one that is also injected as pure kinetic energy but with ejecta radial velocities distributed in the form of a skewed Gaussian function (dark blue; identical to the free-field curve in Fig.~\ref{fig:outflow_cooling} first panel). Result from an adiabatic run is plotted for comparison (grey). The analytical solutions of free-field and semi-confined are shown in blue and red dashed lines respectively.  }
    \label{fig:v_detector_ff}
\end{figure}

We discuss the reasons as to why the free-field velocities at the detector in the cooling run remain steady throughout its evolution (see Fig.~\ref{fig:outflow_cooling}; first panel). Such discrepancy with the analytical model may be explained using several extra test simulations. The results are presented in Fig.~\ref{fig:v_detector_ff}. The dark blue curve is identical to that in Fig.~\ref{fig:outflow_cooling}. 

First, by simply switching off cooling (grey), we recover the analytical solution. Despite the SN is injected as pure kinetic energy, the energy is soon thermalized in the shocks and thus reproduces the adiabatic curve with SN injected as thermal energy, shown in Fig.~\ref{fig:outflow_adiabatic}. This indicates that the peculiar velocity curve in Fig.~\ref{fig:outflow_cooling} is a direct consequence of our cooling implementation. 

Once radiative cooling effects are incorporated, velocities from a SN injected as pure thermal energy (navy) remain similar to the adiabatic case only up to around $0.20\ \mathrm{Myr}$ -- beyond this point, the curve begins to rise. The reason may be attributed to the low gas density in our free-field ambient medium. Fig.~\ref{fig:teq_solutions} shows that gas equilibrium temperatures $T_\mathrm{eq}$ at densities beneath $10^{-25}\ \mathrm{g\ cm^{-3}}$ rise up to $10^4\ \mathrm{K}$, which corresponds to the ISM warm phase, from the fact that cooling is highly inefficient in low-density regions. The gas around the evacuated progenitor drops in density and in turn becomes heated by the background heating constant $\Gamma$ \citep[cf.][]{koyamainutsuka02}. As thermal energy drives gas expansions, this heating causes the rise in velocities towards the later stages.  

The second factor that was responsible for the large velocities at the early stages in Fig.~\ref{fig:outflow_cooling} (dark blue in Fig.~\ref{fig:v_detector_ff}) is the velocity profile of the ejecta particles. In our simulations, we set the radial velocities to take the form of a skewed Gaussian function in order to create a dense shell. However, doing so extends the driving timescale and creates a \textit{series} of shock waves, deviating it from being an energy point source as assumed in the Sedov--Taylor solution. We illustrate our point by performing an identical test run but with a uniform velocity profile assigned to the ejecta particles (light blue). It can be seen that setting up a coherent shock immediately recovers the pure thermal energy case (navy), in which the SN is closest to being a point source explosion. The curves agree well with the analytical model before they are `deflected' by the heating. 

From the above, we argue that the flat velocity curve in Fig.~\ref{fig:outflow_cooling} is only a consequence of our low-density ambient medium and our choice of ejecta velocity distribution. Despite these influences, however, the perturbations induced by a free-field SN are still below that of the semi-confined. It hence provides strong evidence to support the idea that semi-confined SNe can impose a higher dynamical impact locally.


\bsp	
\label{lastpage}
\end{document}